\documentclass{aa}  

\usepackage[utf8]{inputenc} 
\usepackage[T1]{fontenc}    
\usepackage{enumitem}
\usepackage{csquotes}
\usepackage{nicefrac}

\usepackage{graphicx}
\graphicspath{ {./figures/} }

\usepackage{txfonts}
\usepackage[version=4]{mhchem}
\usepackage{siunitx}
\usepackage{tabularx}
\usepackage{xr}
\usepackage{ltablex}
\usepackage{booktabs}
\usepackage{placeins}

\usepackage{hyperref}
\hypersetup{unicode=true, pdftoolbar=true, pdfmenubar=true, pdffitwindow=false, pdfstartview={FitH}, pdftitle={Redox state and interior structure control on the long-term habitability of stagnant-lid planets}, pdfauthor={Philipp Baumeister, Nicola Tosi, Caroline Brachmann, John Lee Grenfell, Lena Noack},	pdfsubject={},	pdfcreator={\LaTeX\ with package \flqq hyperref\frqq}, pdfproducer={pdfTeX \the\pdftexversion.\pdftexrevision}, pdfkeywords={exoplanets, thermal evolution, planet interior, atmosphere, feedback mechanism, outgassing, habitability, water, redox state}, pdfnewwindow=true, colorlinks=true,linkcolor=black,citecolor=black,filecolor=magenta,urlcolor=black, pdfborder= 0 0 0}

\usepackage{cleveref}
\DeclareSIUnit\year{yr}

\newcommand*\subtxt[1]{_{\textnormal{#1}}}
\DeclareRobustCommand\_{\ifmmode\expandafter\subtxt\else\textunderscore\fi}

\newcommand{\overbar}[1]{\mkern 1.5mu\overline{\mkern-1.5mu#1\mkern-1.5mu}\mkern 1.5mu}

\newcommand{\Ra}{\mathit{R\mkern-1.2mu a}}
\newcommand{\St}{\mathit{S\mkern-1.2mu t}}
\newcommand{\dt}[1]{\frac{\mathrm{d} #1}{\mathrm{d} t}}
\newcommand{\del}[2]{\frac{\partial #1}{\partial #2}}
\newcommand{\Int}[4]{\int\limits_{#1}^{#2} \! #3 \, \mathrm{d}#4}
\newcommand{\Me}{M_\oplus}

\begin{document}

   \title{Redox state and interior structure control on the long-term habitability of stagnant-lid planets}

   \subtitle{}

   \titlerunning{Redox state control on the long-term habitability of stagnant-lid planets}
   \authorrunning{Baumeister et al.}

   \author{Philipp Baumeister \inst{1,2}
          \and
          Nicola Tosi \inst{1}
          \and
          Caroline Brachmann \inst{1,3}
          \and
          John Lee Grenfell \inst{1}
          \and
          Lena Noack \inst{3}
          }

   \institute{Institute of Planetary Research, German Aerospace Center (DLR), Rutherfordstraße 2, D-12489 Berlin, Germany\\
              \email{philipp.baumeister@dlr.de}
              \and
              Department of Astronomy and Astrophysics, Technische Universität Berlin, Hardenbergstraße 36, D-10623 Berlin,  Germany
              \and
              Department of Earth Sciences, Freie Universität Berlin, Malteserstr. 74-100, D-12249 Berlin, Germany
              }

   \date{Received December 23, 2022 / Accepted June 7, 2023}

 
  \abstract
   {A major goal in the search for extraterrestrial life is the detection of liquid water on the surface of exoplanets. On terrestrial planets, volcanic outgassing is a significant source of atmospheric and surface water and a major contributor to the long-term evolution of the atmosphere. The rate of volcanism depends on the interior evolution and on numerous feedback processes between atmosphere and interior, which continuously shape atmospheric composition, pressure, and temperature.}
   {
   We explore how key planetary parameters, such as planet mass, interior structure, mantle water content and redox state, shape the formation of atmospheres that permit liquid water on the surface of planets.
   }
   {We present the results of a comprehensive 1D model of the coupled evolution of the interior and atmosphere of rocky exoplanets that combines central feedback processes between these two reservoirs. We carried out more than \num{280000} simulations over a wide range of mantle redox states and volatile content, planetary masses, interior structures and orbital distances in order to robustly assess the emergence, accumulation and preservation of surface water on rocky planets. To establish a conservative baseline of which types of planets can outgas and sustain water on their surface, we focus here on stagnant-lid planets.}
   {We find that only a narrow range of the mantle redox state around the iron-w\"ustite buffer allows the formation of atmospheres that lead to long-term habitable conditions. At oxidizing conditions similar to those of the Earth's mantle, most stagnant-lid planets end up in a hothouse regime akin to Venus due to strong \ce{CO2} outgassing. At more reducing conditions, the amount of outgassed greenhouse gases is often too low to keep surface water from freezing. In addition, Mercury-like planets with large metallic cores are able to sustain habitable conditions at an extended range of orbital distances as a result of lower volcanic activity.}
   {}


   \maketitle
%

\section{Introduction}
The water inventory of a rocky planet originates from the time of formation, with water-bearing materials delivered during accretion onto the protoplanet \citep{obrien2018DeliveryWater, walsh2011LowMass}. Water is brought to the surface and enters the atmosphere during the early magma ocean phase \citep{elkins-tanton2008LinkedMagma,hamano2013EmergenceTwo,lebrun2013ThermalEvolution,nikolaou2019WhatFactors} and later via volcanism throughout the lifetime of the planet \citep{tosi2017HabitabilityStagnantlid,godolt2019HabitabilityStagnantlid}. Water plays an important role in both interior and atmospheric processes. Its presence lowers the melting temperature of rocks \citep{katz2003NewParameterization} and their viscosity \citep{hirth1996WaterOceanic,hirth2004RheologyUpper}, with major implications for global-scale mantle dynamics \citep{nakagawa2015WaterCirculation}, planetary evolution \citep{tosi2017HabitabilityStagnantlid, morschhauser2011CrustalRecycling}, as well as for the emergence of plate tectonics \citep{peslier2017WaterEarth}. In the atmosphere, water is involved in a number of feedback processes controlling the climate, with water vapor strongly contributing to greenhouse heating \citep{kasting1988RunawayMoist, catling2017AtmosphericEvolution}.  Liquid water is also an essential component in carbonate-silicate weathering processes, which help stabilize the climate over geological timescales \citep{walker1981NegativeFeedback}. Furthermore, liquid water is a crucial prerequisite for life \citep{westall2018ImportanceWater, cockell2016HabitabilityReview}. 

Volcanic outgassing links mantle and atmosphere and establishes feedback loops as water and other volatiles are removed from the mantle and brought into the atmosphere. The composition of volcanic gases, in turn, is in large parts determined by the composition and pressure of the atmosphere \citep{gaillard2014TheoreticalFramework}. Planetary mass and interior structure play an additional important role in shaping the rate of volcanism and interior dynamics \citep{noack2014CanInterior, noack2017VolcanismOutgassing, stamenkovic2012INFLUENCEPRESSUREDEPENDENT}. 

Many previous interior-atmosphere studies of the habitability of rocky exoplanets over geological timescales have either considered only selected feedbacks, or investigated only planets with an Earth-like mass and structure \citep{noack2014CanInterior, noack2017VolcanismOutgassing, tosi2017HabitabilityStagnantlid, foley2018CarbonCycling, honing2019CarbonCycling, godolt2019HabitabilityStagnantlid, bower2019LinkingEvolution, kite2020ExoplanetSecondary, spaargaren2020InfluenceBulk, liggins2022GrowthEvolution}. 


Most parameters driving the interior evolution of rocky planets, especially mantle parameters such as the redox state and initial water content, are difficult to constrain and often inaccessible to observation. The large size of the parameter space and the numerous feedbacks between interior and atmosphere require a statistical approach to obtain a thorough overview over which planets are the most likely to exhibit oceans. 
In this work, we present the results of more than \num{280000} coupled interior-atmosphere evolutions of rocky planets with different initial conditions and interior structures (Section \ref{sec:interior_structure}) and investigate the emergence of surface water based on planet mass, age, mantle volatile content and redox state, and orbital distance to the host star. Our model combines a 1D parameterized convection model (Section \ref{sec:1D_model}) to simulate the mantle and core evolution of rocky planets up to \SI{3}{\Me}, as well melting and volcanic outgassing, with a gray atmosphere model tracking the evolution of atmospheric composition, pressure, and temperature. We focus on stagnant-lid planets in order to establish a baseline of worlds that could sustain liquid water on their surface. These are less prone to sustain habitable conditions than plate-tectonics planets due to a strongly reduced recycling of volatiles into the mantle, which would otherwise favor long-term, temperate climates \citep{walker1981NegativeFeedback, kasting1993HabitableZones}. 
We include a comprehensive array of feedback processes: 
\begin{enumerate}
\item A speciation model to self-consistently treat the outgassing of volatile C-O-H species from surface melts \citep{ortenzi2020MantleRedox} (Section \ref{sec:outgassing_model}). The final composition of outgassed volatiles depends primarily on the redox state of the melt and the current composition and pressure of the atmosphere (Section \ref{sec:atmosphere}). At oxidizing conditions, the predominantly outgassed species are \ce{H2O} and \ce{CO2}, while reducing conditions favor the outgassing of oxygen-poorer species, mainly \ce{H2} and \ce{CO}.
\item A simple scheme for surface water accumulation and evaporation (Section \ref{sec:water_cycle}). We assume the atmosphere to be fully saturated in water. Any excess water condenses to form a surface ocean or ice. 
\item A stagnant-lid \ce{CO2} weathering cycle \citep{honing2019CarbonCycling} (Section \ref{sec:weathering}). On Earth, the long-term carbonate-silicate cycle is important for stabilizing the climate over geological time\-scales. This cycle is primarily driven by plate tectonics, which continuously recycles \ce{CO2} into the mantle via subduction. On stagnant-lid planets, the cycle relies on the production of fresh crust through volcanism. In the presence of liquid water, \ce{CO2} can be weathered and buried under subsequent volcanic eruptions \citep{foley2018CarbonCycling,honing2019CarbonCycling}. 
\item Evolution of \ce{H2} in the atmosphere. According to the evolution of the stellar XUV flux (Section \ref{sec:stellar_evolution}), \ce{H2} can be lost due to atmospheric escape (Section \ref{sec:escape}) and replenished by volcanic outgassing, particularly under reducing conditions. Since the surface pressure and atmospheric composition play an important role in the outgassing of \ce{H2O}, a thick (potentially primordial) atmosphere can limit the outgassing of water, especially in the early, active phase of a planet's evolution. In addition, \ce{H2} can act as a greenhouse gas via collision-induced absorption \citep{pierrehumbert2011HydrogenGreenhouse}, which we also take into account (Section \ref{sec:atmosphere}).
\end{enumerate}

\section{Methods}
\subsection{Planet interior structures}
\label{sec:interior_structure}

The diversity in densities of low-mass exoplanets hints at a high degree of variation in interior structures \citep{jontof-hutter2019CompositionalDiversity}. We model rocky planets between 0.5 and 3 Earth masses. While many potentially rocky exoplanets with larger masses have been observed, the large pressure gradient in planets more massive than 5-6 $M_\oplus$ can prevent melt from reaching the surface and thus impede the outgassing of volatiles \citep{noack2017VolcanismOutgassing}. Additionally, the mantle rheology and interior dynamics of massive super-Earths are poorly understood \citep{stamenkovic2011ThermalTransport, karato2011RheologicalStructure}. 
We model the interior structure of each planet with our interior structure code TATOOINE \citep{baumeister2020MachinelearningInference}. Each planet consists of an iron-rich core and a silicate mantle of Earth-like composition. We consider planets with core radius fractions ranging from 0.3 (a "Moon-like" planet) up to 0.7 (a "Mercury-like" planet). From these modeled interior structures, we calculate the average density of the core ($\rho\_c$) and mantle ($\rho\_m$) for use in the parameterized convection model:
\begin{equation}
    \label{eq:rhoC_rhoM}
    \rho\_c = \frac{M\_c}{4/3\pi R\_c^3}, \quad \rho\_m = \frac{M\_p - M\_c}{4/3\pi\left(R\_p^3 - R\_c^3\right)},
\end{equation}
where $M\_p$ and $R\_p$ are the mass and radius of the planet, and $M\_c$ and $R\_c$ are the core mass and radius, respectively. We assume a constant gravitational acceleration $g$ throughout the planetary mantle, with
\begin{equation}
    g = \frac{G M\_p}{R\_p^2},
\end{equation}
where $G$ is the gravitational constant.

Figure \ref{fig:MRcrf} in Appendix \ref{sec:additionl_figs} shows the range of investigated planets in the form of a mass-radius diagram.

\subsection{1D parameterized convection model}
\label{sec:1D_model}

We employ a one-dimensional (1D) parameterized convection model to simulate the thermal evolution of the mantle and core of stagnant-lid rocky planets, as well as the melting of mantle rocks and volcanic outgassing of volatiles \citep{stamenkovic2012INFLUENCEPRESSUREDEPENDENT, grott2011VolcanicOutgassing, tosi2017HabitabilityStagnantlid}.

We focus on stagnant-lid planets, since the emergence of and transition into plate tectonics is still poorly understood, and (especially for exoplanets) the question of which planets are the most likely to have plate tectonics has proven controversial, with many studies giving contradicting results \citep{oneill2016WindowPlate, noack2014PlateTectonics, stein2013InfluenceMantle, vanheck2011PlateTectonics, korenaga2010LikelihoodPlate, valencia2007InevitabilityPlate, oneill2007GeologicalConsequences}. Furthermore, plate tectonics favors establishing stable, temperate climates due to the efficient recycling of volatiles into the mantle \citep{walker1981NegativeFeedback}. On stagnant-lid planets, this recycling is strongly reduced. In this sense, our study provides a conservative baseline to assess whether a planet can sustain liquid water on its surface. In addition, except for Earth, the terrestrial planets of the Solar System are in a stagnant-lid regime at present day, and likely have been for a majority of their evolution \citep{oneill2007ConditionsOnset, tosi2021MercuryMoon} (A direct comparison to present-day Venus can be found in the Appendix \ref{sec:venus}).

Partial melting occurs everywhere where the mantle temperature profile exceeds the solidus. The model accounts for the partitioning of incompatible trace elements (water, \ce{CO2}, and radiogenic elements) between mantle, crust, and, in the case of volatiles, the atmosphere via volcanic outgassing. In addition, the presence of water depresses the solidus temperature \citep{katz2003NewParameterization}. We focus here on fully extrusive volcanism, where all the melt produced reaches the surface and is subject to outgassing. In Sect. \ref{sec:intrusive_volc}, we discuss the impact of intrusive volcanism on our results.
Once melt reaches the surface, dissolved volatiles can be outgassed into the atmosphere. This process is subject to a number of limiting factors, such as the solubility of volatiles in the melt and the evolving atmosphere composition and pressure. We model the composition of outgassed species with a speciation model based on the chemical equilibrium of volatiles between melt and atmosphere (Sect. \ref{sec:outgassing_model}).

We neglect an early magma ocean phase and focus here on the outgassing of water only via volcanism. This provides a conservative estimate of the total amount of water that can be expected to be outgassed. While magma ocean solidification can result in the formation of a thick steam atmosphere, on an Earth-like planet this would rapidly collapse to form an early ocean \citep{elkins-tanton2011FormationEarly,lebrun2013ThermalEvolution}. Additionally, young planets likely experience significant water loss shortly after formation due to high stellar XUV activity \citep{tian2018WaterLoss}. In Section \ref{sec:primordial_co2}, we discuss the influence of early steam and \ce{CO2} atmospheres that may follow magma ocean solidification \citep{lebrun2013ThermalEvolution,nikolaou2019WhatFactors}.


A detailed description of the convection model as well as the melting and outgassing scheme can be found in the Appendices \ref{sec:convection_model} and \ref{sec:melting}.

\subsection{Outgassing speciation model}
\label{sec:outgassing_model}
We adapt the model by \citet{ortenzi2020MantleRedox} to calculate the chemical speciation of volatiles within the C-O-H system during outgassing from surface melts, based on the amount of dissolved \ce{H2O} and \ce{CO2}.
We follow the approach by \citet{holloway1998GraphitemeltEquilibria} and \citet{grott2011VolcanicOutgassing} to calculate the concentration of \ce{CO2} in melts. We assume a sufficiently reduced mantle to allow for carbon to occur as graphite, with an oxygen fugacity $f_{\ce{O2}}$ ranging from -3 to +3 in $\log_{10}$ units above and below the IW buffer. The (depth-dependent) concentration of \ce{CO2} in the melt ($X\_{liq}^{\ce{CO2}}$) is then given by the concentration of carbonate ($X\_{liq}^{\ce{CO3^2-}}$)
\begin{subequations}
    \label{eq:X_liq_CO2}
    \begin{align}
        X\_{liq}^{\ce{CO2}}(r) &= \frac{b X\_{liq}^{\ce{CO3^2-}}\!(r)}{1 + (b-1) X\_{liq}^{\ce{CO3^2-}}\!(r)}\\ 
        X\_{liq}^{\ce{CO3^2-}}\!(r) &= \frac{K\_{II} K\_{I} f_{\ce{O2}}}{1 + K\_{II} K\_{I} f_{\ce{O2}}},
    \end{align}
\end{subequations}
where $K\_{II}$ and $K\_{I}$ are equilibrium constants governing the reaction of forming carbonate and graphite from \ce{CO2}, respectively, and $b$ is a constant. $K\_{II}$, $K\_{I}$, and $b$ are all determined appropriately for Hawaiian basalts \citep{holloway1998GraphitemeltEquilibria}. We calculate the melt concentration of \ce{H2O} ($X\_{liq}^{\ce{H2O}}$) based on a model of fractional melting as described in the Appendix \ref{sec:melting}, assuming a partition coefficient $\delta_{\ce{H2O}}=0.01$.
The solubility of \ce{H2O} and \ce{CO2} is governed by melt-gas equilibrium reactions according to \citet{iacono-marziano2012NewExperimental}:
\begin{subequations}
    \begin{align}
        \ce{H2O^{[fluid]} + O^2-^{[melt]} &<=> 2 OH^-^{[melt]}}\\
        \ce{CO2^{[fluid]} + O^2-^{[melt]} &<=>  CO3^2-^{[melt]}}
    \end{align}
\end{subequations}

We assume that all of the \ce{CO} and \ce{H2} generated are outgassed due to their low solubility in silicate melts. The final molar composition of outgassed species is then governed by the following gas-gas equilibria:
\begin{subequations}
    \begin{align}
        \ce{H2^{[fluid]} + 1/2 O2 &<=> H2O^{[fluid]}}\\
        \ce{CO^{[fluid]} + 1/2 O2 &<=>  CO2^{[fluid]}},
    \end{align}
\end{subequations}
which can be converted to weight fractions $X\_{outg}^i$ based on the molar mass of the magma (see Table \ref{tab:magma_composition} in the Appendix), and ultimately into a mass rate (See Eq. \eqref{eq:dM_outg} in  Appendix \ref{sec:melting}). We do not include \ce{CH4} in the speciation model, which starts to become relevant only at lithospheric pressures and colder temperatures that are not reached here. In none of the models investigated did the outgassed \ce{CH4} concentration exceed \SI{e-11}{ppm}.

The rate of the gas-gas equilibrium reactions depend mainly on the melt temperature and oxygen fugacity \citep[see e.g.][]{ortenzi2020MantleRedox, gaillard2014TheoreticalFramework}, with the oxygen fugacity being dependent on the degassing pressure and temperature as well. We assume that all melt reaches the surface where it is subject to the current atmospheric surface pressure. To determine the melt temperature, we calculate the volume-averaged temperature and pressure of the melt region and obtain the surface melt temperature by moving the melt adiabatically to the surface.


\subsection{Atmosphere model}
\label{sec:atmosphere}

\ce{H2O} and \ce{CO2} are potent greenhouse gases and can strongly modify the surface temperature of a planet. \ce{H2}, while not being a strongly absorbing molecule on its own, can also act as a greenhouse gas through collision-induced absorption \citep{pierrehumbert2011HydrogenGreenhouse}, which can be especially relevant for planets with hydrogen-dominated atmospheres. To estimate the surface temperature for a wide range of atmospheric compositions, we adopt a simple two-stream radiative gray atmosphere model to calculate greenhouse heating at the surface \citep{catling2017AtmosphericEvolution}. The surface temperature can be expressed in terms of the optical depth $\tau$ of the atmosphere and the equilibrium temperature $T\_{eq}$ of the planet:
\begin{equation}
    T\_s = T\_{eq} \left( 1 + \frac{\tau}{2}\right)^{\nicefrac{1}{4}} \quad \text{with} \hspace{0.6em} T\_{eq} = \left(\frac{(1-A) S_\odot}{4 \sigma} \right)^{\nicefrac{1}{4}},
\end{equation}
where $S_\odot$ is the solar insolation at the top of the atmosphere, $A$ is the bond albedo, and $\sigma$ is the Stefan-Boltzmann constant. We assume an Earth-like albedo of $0.3$ for all planets considered in this study.
Following \citet{abe1985FormationImpactgenerated} and \citet{pujol2003AnalyticalInvestigation}, the optical depth of the atmosphere is given by 
\begin{equation}
    \tau = \sum_i{\tau_i} = \sum_i{\frac{3 k'_i P_i}{2 g}},
\end{equation}
where $g$ is the planet gravity, $P_i$ is the partial pressure of a given atmosphere species $i$, and $k'_i$ is the extinction coefficient relative to this pressure. $k'_i$ can be expressed using the extinction coefficient $k_{0,i}$ at standard atmospheric pressure $P_0$
\begin{equation}
    k'_i = \left( \frac{k_{0,i} g}{3 P_0}\right)^{\nicefrac{1}{2}}.
\end{equation}

In order to approximate the collision-induced absorption of \ce{H2}, we choose a value of $k_{0,\ce{H2}}=\SI{2e-2}{\m\squared\per\kg}$, which fits well to the results of \citet{pierrehumbert2011HydrogenGreenhouse} (see also Fig. \ref{fig:pierrehumbert} in the Appendix). A validation of our atmosphere model against a 3D climate model can be found in \citet{honing2021EarlyHabitability}.

\subsection{Water condensation and evaporation}
\label{sec:water_cycle}
We assume the atmosphere to be fully saturated in \ce{H2O}, with any excess outgassed water condensing into a surface ocean or forming ice. This provides an upper limit for the mass of water in the atmosphere, and consequently also for the contribution of water to greenhouse heating. We calculate the saturated partial pressure (in Pa) of \ce{H2O} from the saturation vapor pressure curve by \citet{alduchov1996ImprovedMagnus}: 
\begin{equation}
    \label{eq:vapor_pressure}
    P\_{vapor} = 610.94 \exp\left( \frac{17.625\:(T - 273.15) }{T - 30.1}\right) \quad \text{for } T \geq \SI{273.15}{K},
\end{equation}
where $T$ is the temperature in \si{K}. If the surface temperature drops below the freezing point of water, we assume that the surface of an existing ocean would freeze. In this case, we use a vapor pressure curve from \citet{alduchov1996ImprovedMagnus} defined over a plane of ice:
\begin{equation}
    \label{eq:vapor_pressure_ice}
    P^{\text{ice}}\_{vapor} = 611.21 \exp\left( \frac{22.587\:(T - 273.15) }{T + 0.71}\right)\quad \text{for } T< \SI{273.15}{K}.
\end{equation}

The combination of water evaporation and greenhouse heating from water acts as a positive feedback process which can give rise to a runaway process commonly described by the runaway greenhouse scenario \citep{komabayasi1967DiscreteEquilibrium,ingersoll1969RunawayGreenhouse,kasting1988RunawayMoist,nakajima1992StudyRunaway}. While with this model we do observe these rapid transitions from temperate conditions to hot, dense atmospheres, caused by the runaway evaporation of water, we are not performing a full, rigorous description of the runaway greenhouse. To avoid ambiguities and to make clear that our process is not a complete representation of the runaway greenhouse scenario, we are calling these "hothouse" states here.

\subsection{Carbon weathering cycle}
\label{sec:weathering}
On Earth, the long-term carbon-silicate cycle is an important process to stabilize the climate over geological time-scales \citep{walker1981NegativeFeedback}. This cycle is primarily driven by plate tectonics, where \ce{CO2} can be continuously recycled into the mantle with subducting plates and fresh crust constantly produced at mid-ocean ridges. Stagnant-lid planets, on the other hand, do not have subducting plates. A carbon cycle on these planets relies on the continuous production of new crust through hot-spot volcanism, which can be weathered in the presence of liquid water. The carbonated crust may then be buried by subsequent volcanic eruptions, sinking downward in the mantle. The rate of weathering is therefore closely coupled to the crust production rate. We follow the model by \citet{honing2019CarbonCycling}, assuming that the rate of \ce{CO2} weathering depends on the partial pressure of \ce{CO2} in the atmosphere. We additionally assume that all newly formed crust is subject to weathering. The weathering rate $\Phi\_w$ can then be expressed as a function of the crustal growth rate $\dt{M\_{cr}}$ and the partial \ce{CO2} pressure in the atmosphere $P_{\ce{CO2}}$, and scaled to the seafloor weathering rate on Earth \citep[for more details, see][]{honing2019CarbonCycling}:
\begin{equation}
    \label{eq:weathersl}
    \Phi\_w=\frac{X\_E \xi\_E}{f\_E \phi_E}\left(\dt{M\_{cr}}\right)\left(\frac{P_{\ce{CO2}}}{P_{\ce{CO2},\text{E}}}\right)^\alpha,
\end{equation}
where $\alpha=0.23$ is a scaling exponent, and $P_{\ce{CO2},\text{E}} = \SI{4e-4}{bar}$ is the present-day partial pressure of \ce{CO2} in Earth's atmosphere. The other parameters are factors scaling the weathering rate to the observed present-day seafloor weathering rate on Earth: $X\_e$ is the present-day Earth mid-ocean ridge \ce{CO2} concentration in the melt, $\xi\_E$ is the proportion of seafloor weathering to the total weathering rate on Earth, $f\_E$ is the present-day fraction of carbonates that are recycled back into Earth's mantle, and $\phi_E$ is the fraction of carbonates that remain stable during subduction.

Carbonates are stable only up to a certain pressure-dependent temperature. Once the sinking carbonated crust reaches this temperature, it undergoes decarbonation and releases its \ce{CO2}, which will rise through cracks in the crust and eventually return to the atmosphere \citep{foley2018CarbonCycling, honing2019CarbonCycling}. This means that, in contrast to plate tectonics, \ce{CO2} is generally not recycled back into the mantle in the stagnant lid regime. We calculate the depth $z\_{decarb}$ at which decarbonation occurs following \citet{honing2019CarbonCycling}: 
\begin{equation}
    z\_{decarb}=\frac{T\_s-B\_{decarb}}{A\_{decarb}-\displaystyle\frac{T\_m-T\_s}{D\_l+d\_u }},
    \label{eq:zdecarbsl}
\end{equation}
where $A\_{decarb}=\SI{3.125e-3}{\K\per\m}$ and $B\_{decarb}=\SI{835.5}{K}$ are constants related to the decarbonation temperature \citep{foley2018CarbonCycling}, $T\_m$ is the mantle temperature, $D\_l$ is the thickness of the stagnant lid, and $d\_u$ the thickness of the upper thermal boundary layer of the convecting mantle.

During the evolution of the planet, our model continuously tracks the depth of the previously weathered crustal layers. Once the decarbonation depth is reached, the carbon content of the layer is released as \ce{CO2}. To avoid numerical instabilities, the released \ce{CO2} is first stored in a temporary volatile buffer, which releases 10\% of its \ce{CO2} content into the atmosphere at every time step \citep[see also][]{honing2019CarbonCycling}.

Carbon weathering requires the presence of liquid water. Therefore, we set the weathering rate to zero if either no surface water is present, or if the surface temperature lies below the freezing point of water. For the latter, this means that any existing ocean in our model will freeze over so that little exchange of \ce{CO2} with the ocean is possible. Nevertheless, in both cases, the burying of previously formed carbonates and decarbonation continue as long as there is active volcanism.

\subsection{Stellar evolution}
\label{sec:stellar_evolution}
We focus on planets around G-type stars with one solar mass. We account for an increasing stellar insolation $ S_\odot$ over the lifetime of the host star by using the parameterization by \citet{gough1981SolarInterior},
\begin{equation}
    S_\odot(t) = S_{\odot, 0} \left(1 + \frac{2}{5} \left(1 - \frac{t}{t_0} \right) \right)^{-1},
\end{equation}
where $S_{\odot, 0}$ is the insolation at the planet at present day $t_0 = \SI{4.5}{Gyrs}$.

In order to model atmospheric escape processes, we follow \citet{owen2017EvaporationValley} for a parametrization of the stellar XUV flux evolution:
\begin{equation}
    F\_{XUV}(t) = 
    \begin{cases} 
      F\_{sat} & \mbox{for } t<t\_{sat} \\
      F\_{sat}\left( \displaystyle \frac{t}{t\_{sat}}\right)^{-1.5} & \mbox{for } t\geq t\_{sat} 
   \end{cases}
   \qquad \text{with } F\_{sat} = 10^{-3.5} S_{\odot, 0},
\end{equation}
with a saturation timescale of $t\_{sat} = \SI{100}{Myrs}$.

\subsection{Atmospheric escape}
\label{sec:escape}
To model the transition from a primary \ce{H2} to a secondary outgassed atmosphere and to treat the loss of later outgassed \ce{H2}, we consider hydrodynamic escape of \ce{H2}. For hydrogen-dominated atmospheres, the maximum rate at which hydrogen can escape is limited by the amount of energy from XUV radiation that the atmosphere can absorb. The energy-limited mass-loss rate is given by
\begin{equation}
    \label{eq:flux_el}
    \dot{M}\_{el} = \frac{\varepsilon \pi R\_p R\_{atm}^2 F\_{XUV}}{G M\_p},
\end{equation}
where $\varepsilon$ is an efficiency factor we here take to be 0.15 following \citet{kite2020ExoplanetSecondary}, and $R\_{atm}$ is the planet radius at the top of the atmosphere, which we define at \SI{20}{mbar}.

In the case of hydrogen existing as a minor atmospheric component within a background of heavier species, the loss of hydrogen is limited by the rate at which it can be supplied from the lower parts of the atmosphere. This diffusion-limited escape provides an upper limit to hydrodynamic escape. The diffusion-limited mass loss rate can be expressed as
\begin{equation}
    \label{eq:flux_dl}
    \dot{M}\_{dl} = 4 \pi R\_{atm}^2 \frac{m_{\ce{H2}}}{N\_A} b_{aj} \chi_{\ce{H2}} \left( \frac{1}{H_a} - \frac{1}{H_{\ce{H2}}}\right),
\end{equation}
where $m_{\ce{H2}}$ is the molar mass of molecular hydrogen, $N\_A$ is Avogadro's number, and $\chi_{\ce{H2}}$ is the molar mixing ratio of hydrogen in the atmosphere. $H_{\ce{H2}}$ and $H_a$ are the unperturbed scale heights of \ce{H2} and the background gas, respectively. $b_{aj}$ is the binary diffusion coefficient between the escaping \ce{H2} and the heavier background gas. In our case, the background gas consists of varying amounts of \ce{CO2}, \ce{CO}, and \ce{H2O}. We calculate $b_{aj}$ as the sum of the respective binary diffusion coefficients $b_{\ce{CO2}}=\SI{3e21}{\per\m\per\s}$, $b_{\ce{CO}}=\SI{3e21}{\per\m\per\s}$, and $b_{\ce{H2O}}=\SI{4.3e21}{\per\m\per\s}$ between \ce{H2} and the background gases \ce{CO2}, \ce{CO}, and \ce{H2O}, weighted by the respective relative volume mixing ratios (VMR) $\chi_{\ce{CO2}}$, $\chi_{\ce{CO}}$, and $\chi_{\ce{H2O}}$ (See also Table \ref{tab:parameters} in the Appendix for references):
\begin{equation}
    b_{aj} = b_{\ce{CO2}} \chi_{\ce{CO2}} + b_{\ce{CO}} \chi_{\ce{CO}} + b_{\ce{H2O}} \chi_{\ce{H2O}}.
\end{equation}

The transition from energy-limited to diffusion-limited escape, and thus from a hydrogen-dominated to a secondary atmosphere, is currently not well understood and requires the use of detailed hydrodynamical models \citep{owen2019AtmosphericEscape, zahnle2019StrangeMessenger} which are out of the scope of this work, especially since volcanic outgassing continuously changes the atmospheric composition, which can make the hydrodynamical treatment challenging. Here, we opt for smoothly interpolating between energy-limited and diffusion-limited mass loss rates for intermediate \ce{H2} fractions, with the interpolated mass loss rate given by
\begin{equation}
    \label{eq:atm_loss_rate}
    \dot{M}\_{loss} = f\_{el} \dot{M}\_{el} + (1 - f\_{el}) \dot{M}\_{dl},
\end{equation}
where the contribution of energy-limited escape $f\_{el}$ is given by a logistic function \citep[see also e.g.][]{kite2020ExoplanetSecondary}:
\begin{equation}
    \label{eq:el_fraction}
    f\_{el}(\chi_{\ce{H2}}) = \left(1 + \exp\left( -\frac{\chi_{\ce{H2}} - \chi_0}{w}\right)\right)^{-1},
\end{equation}
centered at a hydrogen VMR of $\chi_0=0.15$ and a horizontal scaling $w=0.01$. These parameters are chosen to account for a transition from purely energy-limited escape starting at a hydrogen VMR of 20\% to a purely diffusion-limited escape at a mixing ratio of 10\%.

We do not consider the photodissociation of water in the upper atmosphere and the subsequent loss of hydrogen. Significant water loss will occur once large amounts of water reach the stratosphere, which mainly occurs in planets undergoing a runaway greenhouse regime. On habitable planets, which are the main focus of this study, the tropopause "cold trap" prevents significant amounts of water vapor from reaching the stratosphere \citep{catling2017AtmosphericEvolution}.

\subsection{Investigated parameters and initial conditions}
\label{sec:initial_parameters}
We adopt a Monte-Carlo sampling approach to model an entire population of planets where the initial conditions for each planet are set to random values within given ranges (i.e. with a uniform distribution).

We compute the thermal evolution for a set of $\approx 280\:000$ initial conditions. Each evolution is run up to \SI{8}{Gyrs} to cover a wide range of potentially observable planets. We stop the evolution earlier if the surface temperature exceeds \SI{1500}{K}, at which point the surface rocks would be close to melting. In order to simulate the observation of planets with different ages, we select snapshots of the evolution at up to five randomly chosen times after \SI{100}{Myrs}. For models which finished earlier (because the surface temperature has risen too high), we select fewer snapshots accordingly to ensure a balanced sample of planet ages. This results in a final data set size of $\approx 1\:000\:000$ planets. 

We vary the initial water concentration in the mantle $X^{\ce{H2O}}\_{m,0}$ between 100 and \SI{1000}{ppm}, corresponding to relatively dry and wet conditions, respectively. We allow the mantle oxygen fugacity to vary between three $\log_{10}$ units below and above the iron-wüstite buffer (IW).

Each planet in the parameter study is placed at a fixed distance to its (Sun-like) host star, ranging from a Venus-like orbit (\SI{0.723}{au}) to a Mars-like orbit (\SI{1.524}{au}). In addition, we set the mass of each planet between 0.5 and \SI{3}{M_\oplus} and allow for varied interior structures, where the radius of the core can vary between 30 and 70\% of the planet radius (Section \ref{sec:interior_structure}).
We fix the initial mantle temperature $T\_{m,0}$ at \SI{1700}{K} and prescribe an initial temperature jump of \SI{200}{K} at the CMB.

The main model parameters used in our study are given in Table \ref{tab:parameters} in the Appendix.

We add a note on data presented in Figs. \labelcref{fig:earth_4panels,fig:time_orbit,fig:orbit_crf,fig:sensitivity}. In these figures, we show the surface state of water as a function of initial conditions (oxygen fugacity and initial mantle water content) for points from the data set described above. However, as multiple data points with different ages plot onto the same set of initial conditions, there is a risk to be biased towards later ages. To avoid this, we shuffle the data set before plotting the points. We take advantage of the multiple data points per initial condition to map out regions of long-term behavior: We use a k-nearest neigbor algorithm on every point in the parameter space and color the background according to the prevailing water phase at the surface that is most common among neighboring points.

\section{Results}

\subsection{Characteristic planet evolutions}
\label{sec:example_evolution}

\begin{figure*}[ht!]
    \centering
    \includegraphics[width=\linewidth]{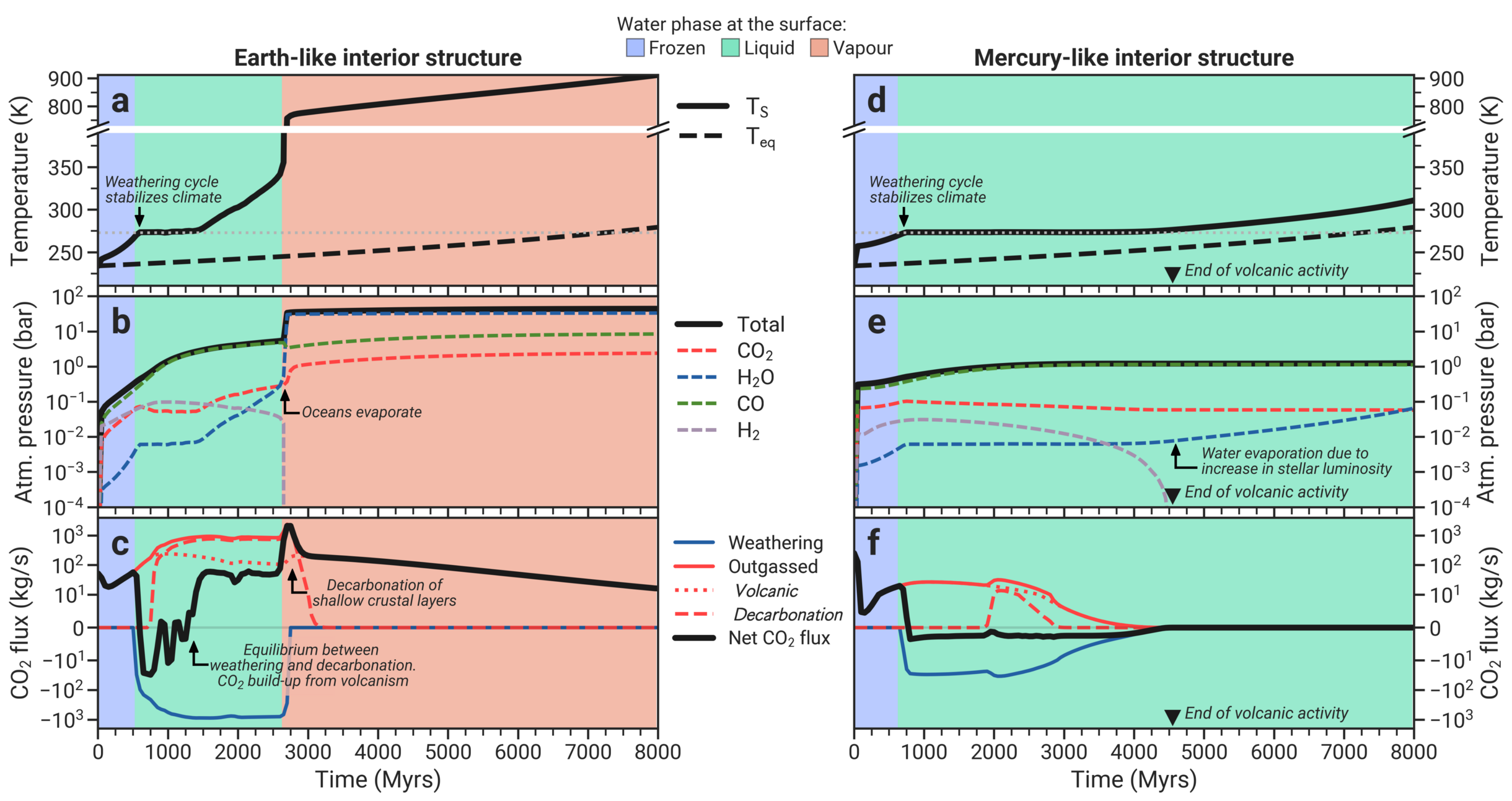}
    \caption{Characteristic evolutions of the atmospheric temperature and pressure and \ce{CO2} flux in and out of the atmosphere for an Earth-mass planet at \SI{1}{au} with an Earth-like (a, b, c) or a Mercury-like (d, e, f) interior structure. The colored background marks the state of water at the planet's surface. Both planets start with the same mantle water content of \SI{250}{ppm} and an oxygen fugacity of 0.05 $\log_{10}$ units above the IW buffer. Panels (a, b, c) and (d, e, f) correspond respectively to the points 1 and 2 marked in Figs. \ref{fig:time_orbit} and \ref{fig:orbit_crf}.}
    \label{fig:evolution}
\end{figure*}

Figure \ref{fig:evolution} demonstrates the characteristic atmospheric evolution for two planets with one Earth mass and either an Earth-like interior structure (Figs. \ref{fig:evolution}a and \ref{fig:evolution}b) or a large, Mercury-like core which makes up 70\% of the interior (Figs. \ref{fig:evolution}c and \ref{fig:evolution}d). Both planets are located at \SI{1}{au} and the initial parameters are the same for both ($\log f\_{\ce{O2}}=-0.05\: \text{IW}$, $X^{\ce{H2O}}\_{m,0}=\SI{250}{ppm}$). These parameters are in the range which enables prolonged habitable conditions for the Mercury-like planet, but puts the atmosphere of the Earth-like planet into a hothouse regime.

In both cases, an outgassed atmosphere of 0.1--\SI{1}{bar} is quickly built up within the first hundred million years. Due to the relatively reducing conditions of the mantle, this initial atmosphere consists mainly of \ce{CO}, \ce{H2}, and \ce{CO2} (\ref{fig:evolution}b and \ref{fig:evolution}d). The surface temperature is initially too low to allow for liquid water (Figs. \ref{fig:evolution}a and \ref{fig:evolution}c), but rises quickly due to the accumulation of greenhouse gases. Once the surface temperature exceeds the freezing point of water, the carbonate-silicate weathering cycle becomes active and is sufficiently strong to counteract the rate of \ce{CO2} outgassing. This keeps the surface temperature just above the freezing point of water. The level of atmospheric \ce{H2} remains relatively stable due to continuous outgassing supply and simultaneous loss from the top of the atmosphere.

Stagnant-lid planets lack an efficient long-term \ce{CO2} sink. While the weathering cycle can temporarily remove \ce{CO2} from the atmosphere, the slow sinking of carbonated crust causes decarbonation within the lid \citep{honing2019CarbonCycling} and thus prevents recycling of carbonates into the mantle. As such, the combined mass of \ce{CO2} in the crust and atmosphere reservoirs continuously grows through the influx of \ce{CO2} from the mantle via volcanic outgassing. Besides the surface temperature, the mantle temperature and lid thickness determine the decarbonation depth (Eq. \ref{eq:zdecarbsl}). This gives rise to two scenarios for the evolution of the atmospheric \ce{CO2} reservoir: 

If the mantle is sufficiently hot, the crust continuously grows, burying carbonated layers. At some point, the bottom of the carbonated crust reaches the decarbonation depth and previously buried \ce{CO2} degasses back into the atmosphere \citep[see Section \ref{sec:weathering} and][]{honing2019CarbonCycling}. For the planet with Earth-like interior, this happens at around \SI{0.8}{\giga\year} (Fig. \ref{fig:evolution}c). Decarbonation degassing quickly becomes the dominant source of atmospheric \ce{CO2}. A steady-state is reached between \ce{CO2} ingassing caused by weathering of fresh rocks and \ce{CO2} degassing from decarbonation of old carbonated layers at the bottom of the crust, as both are tightly coupled to the crustal growth rate and thus to the rate of melt production. As the planet cools, volcanic activity slowly subsides. This reduces the weathering rate, which is tightly coupled to the crust production rate. At around \SI{1.5}{\giga\year}, crust production alone can not sustain the weathering rates necessary to remove all outgassed \ce{CO2}. As the weathering rate is also proportional to the partial pressure of \ce{CO2} (Eq. \ref{eq:weathersl}), the amount of \ce{CO2} in the atmosphere starts to grow until the weathering rate can counteract the increased \ce{CO2} flux into the atmosphere (see also Fig. \ref{fig:supplylimit}). As long as decarbonation takes place, this renders \ce{CO2} weathering largely ineffective as a climate control mechanism. Although decarbonation is closely linked to the crust production rate, the decarbonation flux remains relatively constant, even though the rates of volcanism and volcanic outgassing decrease. This is the result of a hysteresis mechanism in the carbon cycle: Carbonated crustal layers reach the decarbonation depth around 200-300~Myrs after being weathered at the surface (see Fig. \ref{fig:age_decarb} in Appendix \ref{sec:additionl_figs}). The carbonated crust therefore retains a ``memory'' of past \ce{CO2} weathering rates, when the rate of crust production was stronger. Upon decarbonation, relatively high amounts of \ce{CO2} are returned to the atmosphere. The decarbonation flux is thus buffered against a decrease in crust production. In addition, the degassed \ce{CO2} can be weathered again. This establishes a feedback loop, which stabilizes the decarbonation flux over geological timescales.

\ce{CO2} starts accumulating in the atmosphere, driving the surface temperature upwards over geological timescales (in addition to the increasing luminosity of the star), which in turn allows more water to enter the atmosphere. After a \SI{2}{\giga\year}-long habitable phase, the atmosphere transitions into a hothouse state, and the entire water reservoir evaporates to form a thick steam atmosphere with a surface pressure of around \SI{45}{bar}. Once the surface heats up, decarbonation occurs in shallower and shallower layers. This positive feedback leads to a runaway decarbonation, where the entire crustal \ce{CO2} reservoir is rapidly degassed back into the atmosphere (Fig. \ref{fig:evolution}c). At this point, the evolution of this planet does not exhibit any qualitative change, since we do not consider water loss from steam atmospheres. The pressure is too high for much additional \ce{CO2} to be outgassed, so little atmospheric evolution is possible at this point. Some part of the steam atmosphere would be lost via photodissociation, which would allow for more \ce{CO2} outgassing. Such a planet would likely end up with a thick, Venus-like \ce{CO2} atmosphere.

The evolution of the atmosphere is different in the case of the Mercury-like planet. The steep pressure gradient due to the higher gravity (compared to the Earth-like case) allows less melt to rise to the surface, resulting in less overall volcanic activity \citep{noack2017VolcanismOutgassing}. Therefore the crust grows more slowly. In addition, the thinner mantle cools faster. As a result, the growth of the decarbonation depth outpaces the bottom of the carbonated crust for most of the evolution of the planet, with very little decarbonation taking place throughout (Fig. \ref{fig:evolution}f). Here, volcanic outgassing is the dominant source of \ce{CO2}. In contrast to the Earth-like planet, the crust acts as a long-term carbon reservoir, able to store any excess outgassed \ce{CO2}. Volcanism stops completely at around \SI{4.5}{\giga\year} as the core and mantle have cooled to temperatures that no longer make melting possible. Without volcanism, no volatiles are outgassed into the atmosphere, and the remaining \ce{H2} is quickly lost due to atmospheric escape. Likewise, no \ce{CO2} is removed by weathering since this depends on fresh basaltic rock delivered by volcanism, and no decarbonation occurs as the mantle is too cool and, due the lack of crustal growth, carbonated layers are not transported toward the mantle.
Over time, more water vapor enters into the atmosphere as a result of the rising surface temperature due to the increasing luminosity of the star. However, even at \SI{8}{\giga\year}, the planet remains habitable.

\subsection{Role of redox state on habitability}

\begin{figure*}[ht!]
    \centering
    \includegraphics[width=0.8\linewidth]{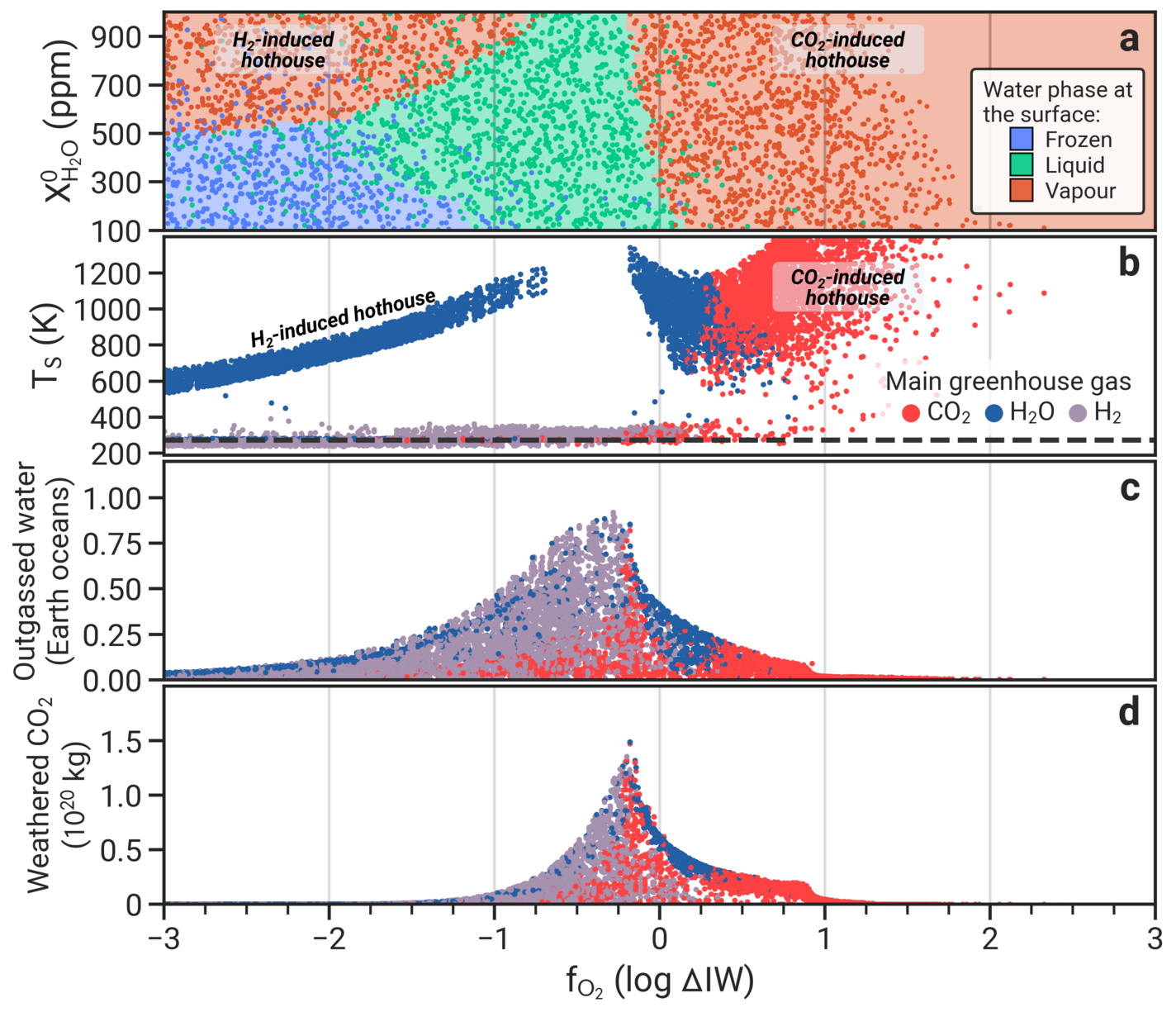}
    \caption{States of a stagnant-lid Earth as a function of the mantle oxygen fugacity. Each point represents a snapshot of the planetary evolution at a randomly selected planet age between 100 Myr and 8 Gyr. The colored points in panel a show the prevailing water phase at the surface as a function of the oxygen fugacity ($f_\mathrm{O_2}$) and initial water content of the mantle ($X^0_\mathrm{H_2O}$). The color background shows the area where the majority of neighboring data points share the same surface conditions. In panels b--d, points are colored according to the greenhouse gas that contributes most to surface heating. Panel b shows the surface temperature, panel c shows the total mass of outgassed water, and panel d shows the total mass of \ce{CO2} removed from the atmosphere via weathering.}
    \label{fig:earth_4panels}
\end{figure*}

In this section, we focus on Earth-mass planets with Earth-like interior structures orbiting a Sun-like star at \SI{1}{au}, with the effects of orbital distance, interior structure and planet being detailed in Section \ref{sec:interior_orbit_mass}. 

We find that the mantle redox state and initial water content are the two main factors limiting the emergence of habitable conditions. Only a narrow range of these two parameters yields long-term stable habitable conditions (Fig. \ref{fig:earth_4panels}a). We can identify well-defined populations of planets with characteristic evolutions of surface habitability. At oxidizing conditions above the iron-w\"ustite (IW) buffer, the majority of planets are in a Venus-like hothouse regime with surface temperatures exceeding \SI{400}{K} (Fig. \ref{fig:earth_4panels}b). At reducing conditions, one to two log$_{10}$ units below the IW buffer, most planets with dry mantles have surface temperatures below the freezing point of water, whereas planets with wet mantles (initial mantle water content exceeding $\sim$500 ppm) are in a hothouse state related to strong \ce{H2} outgassing. Only over a narrow range of mildly reduced mantles are conditions suitable for liquid water to be stable on the surface over long time spans.

The evolution of the surface temperature and the partial pressure of water in the atmosphere are the main factors determining whether water can be outgassed and remain liquid on a planet's surface. The partial pressure of water in turn strongly depends on temperature. At cooler temperatures, most of the outgassed water is in the form of oceans or ice, with relatively small amounts in the atmosphere. At higher temperatures, more and more water vapor can be maintained in the atmosphere. Since water is a strong greenhouse gas, this acts as a positive feedback, where rising surface temperatures cause more water to evaporate \citep{catling2017AtmosphericEvolution}. A rise in surface temperature can be caused by an increase in \ce{CO2} or \ce{H2} and, on longer timescales, an increase in solar luminosity. The outgassing of \ce{CO2} and \ce{H2} therefore predominantly shape the evolution of the surface temperature.

Outgassing rates depend on the amount of melting in the interior. Water in the mantle decreases melting temperatures and viscosity, thus wetter mantles experience more melting and cause more volcanic activity. The oxygen fugacity shapes the composition of outgassed species. At oxygen fugacities above the IW buffer, the main outgassed species are \ce{CO2} and \ce{H2O}. At reducing mantle conditions, on the other hand, outgassing is dominated by the oxygen-poor species \ce{H2} and \ce{CO}. Increasing amounts of either \ce{CO2} or \ce{H2} in the atmosphere eventually push the planet into a hothouse regime, where any existing surface water quickly evaporates to form a \ce{H2O}-rich atmosphere, preventing further water outgassing. 

\begin{figure*}[ht!]
    \centering
    \includegraphics[width=\linewidth]{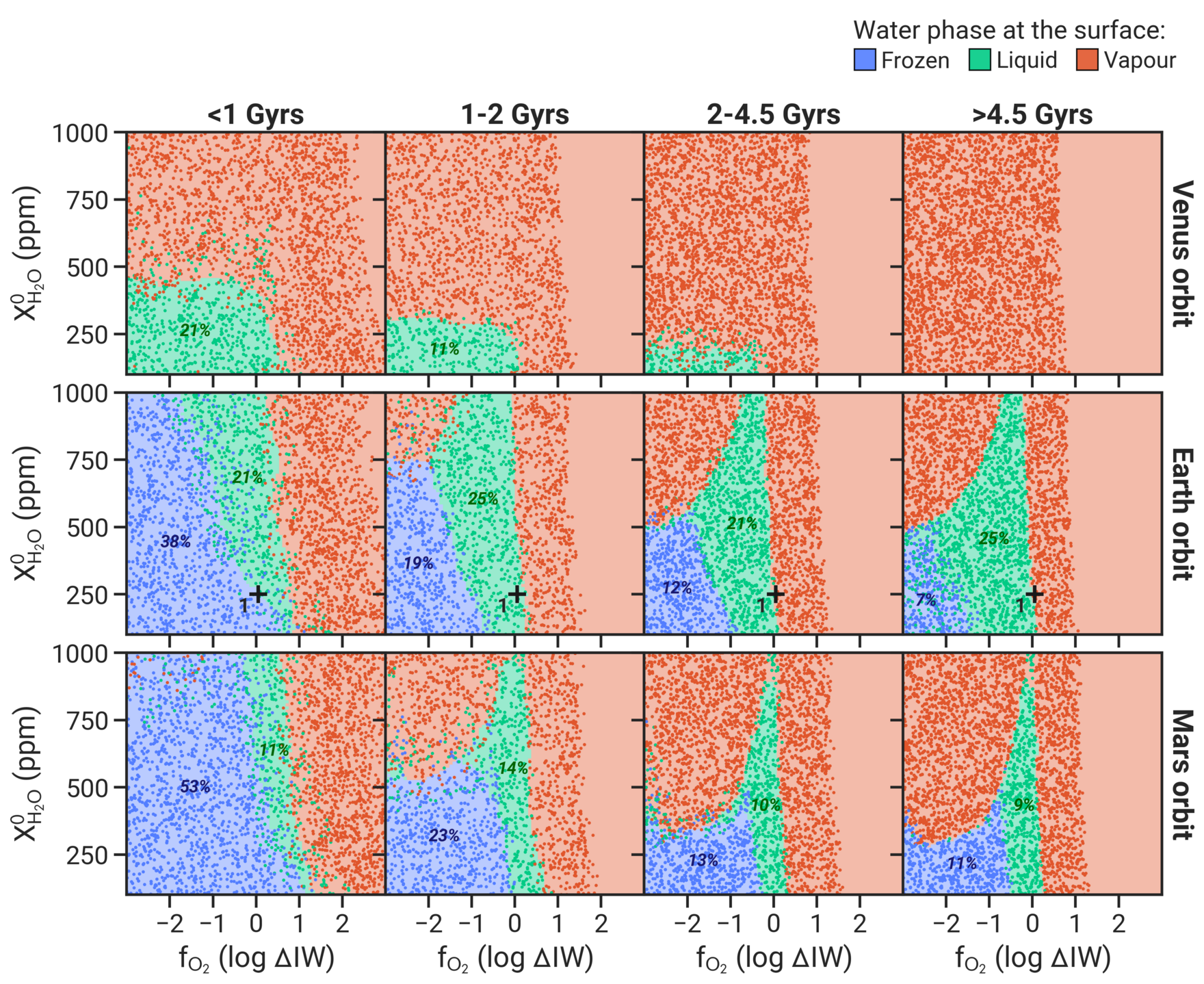}
    \caption{Evolution of habitable conditions of a stagnant-lid Earth at different orbital distances. Each row depicts the evolution of surface conditions of a planet with Earth mass and core size at the orbit of Venus (\SI{0.723}{au}), Earth (\SI{1}{au}) and Mars (\SI{1.524}{au}), respectively. Each plot shows the prevailing surface conditions for water as a function of the oxygen fugacity and initial water content of the mantle, with the color background showing the area where the majority of neighboring data points share the same surface conditions. Similar to Fig. \ref{fig:earth_4panels}a, each point represents a snapshot of the  evolution at a randomly selected planet age within the given age range. The marker in the middle row marks the evolution shown in detail in Fig. \ref{fig:evolution}a,b,c.}
    \label{fig:time_orbit}
\end{figure*}

Strong \ce{CO2} outgassing dominates on planets with an oxygen fugacity more than one log$_{10}$ unit above the IW buffer, which enter a hothouse state within the first billion year of their evolution (Fig. \ref{fig:time_orbit}). Planets closer to the IW buffer may exhibit a short habitable phase.
\ce{H2} becomes the dominant outgassed greenhouse gas at an oxygen fugacity around one log$_{10}$ unit below the IW buffer (see e.g. Fig. \ref{fig:coh} in the Appendix and \citet{ortenzi2020MantleRedox}). \ce{H2} in the atmosphere is steadily lost due to atmospheric escape. Its presence in the atmosphere is therefore maintained only by a continuous replenishment from volcanic outgassing (Fig. \ref{fig:evolution}b), and \ce{H2} only builds up in the atmosphere if the rate of outgassing outweighs the rate of escape. We find that planets with significant \ce{H2} outgassing may enter a hothouse state as well, specifically those with water-rich mantles for which outgassing rates of \ce{H2} are high (Fig. \ref{fig:earth_4panels}a-b, Fig. \ref{fig:time_orbit}). \ce{H2} can build up faster on planets at high orbital distances, where atmospheric loss is less severe. These planets are more likely to end up in a \ce{H2}-induced hothouse as a result (Fig. \ref{fig:time_orbit}).
Habitable conditions occur in the transition regime between \ce{CO2}- and \ce{H2}-dominated outgassing. Here, the combination of \ce{CO2} and \ce{H2} keeps the surface temperature above the freezing point of water, but a hothouse is prevented by weathering of excess \ce{CO2} (Fig. \ref{fig:earth_4panels}d) and continuous loss of \ce{H2} from the top of the atmosphere. These mechanisms favor a stabilization of the climate. Significant amounts of water can be outgassed under these conditions (Fig. \ref{fig:earth_4panels}c). Planets with very reduced, dry mantles can sustain habitable surface conditions for several billion years even at the orbit of Venus, albeit with very thin atmospheres.

\subsection{Effect of interior structure and planetary mass}
\label{sec:interior_orbit_mass}
\begin{figure}[ht!]
    \centering
    \includegraphics[width=0.8\linewidth]{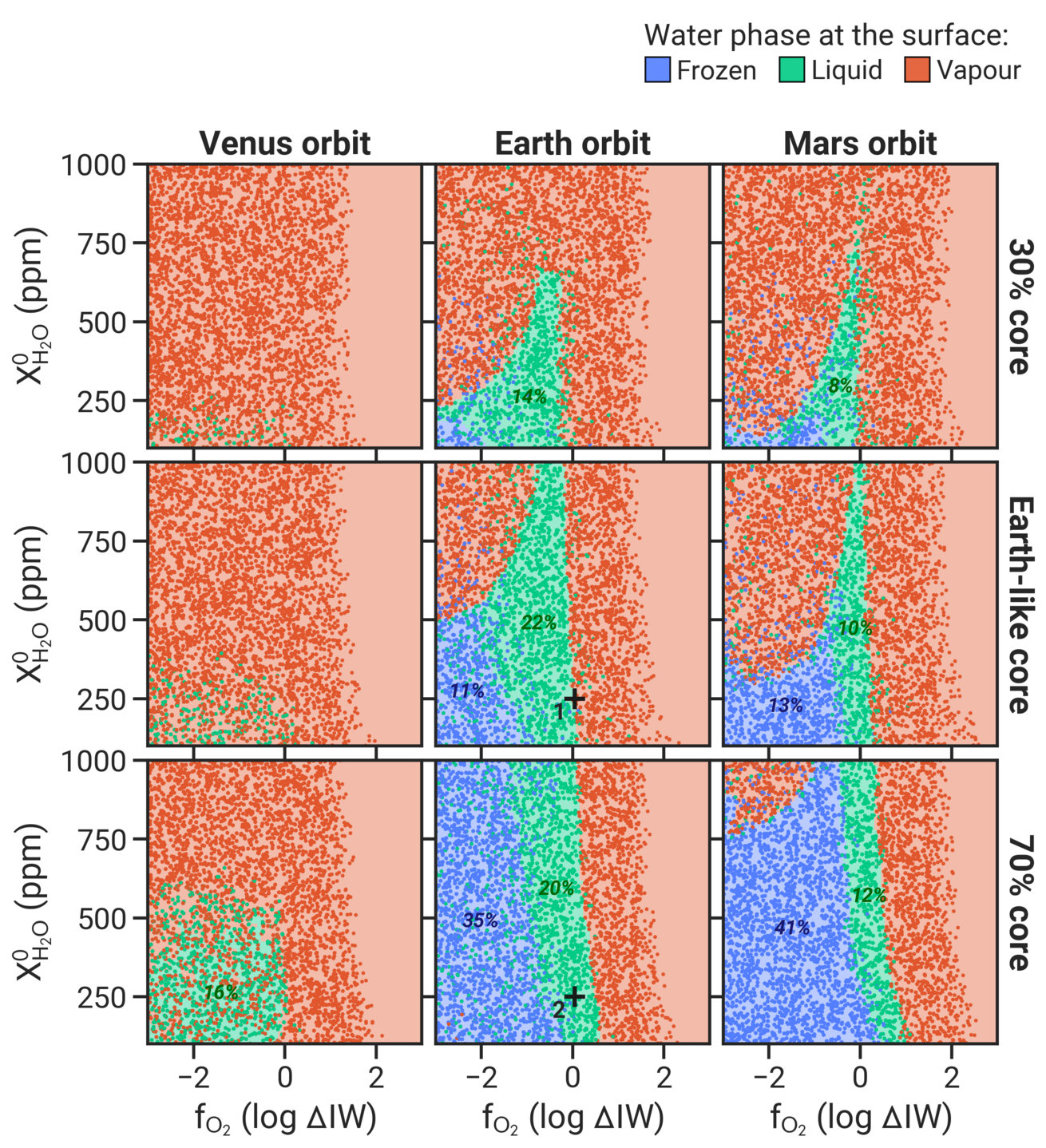}
    \caption{As Fig. \ref{fig:time_orbit}, but showing the prevailing surface conditions of water for different core sizes at different orbital distances. The two markers depict the initial conditions for the detailed evolutions in Fig. \ref{fig:evolution}a, b, c (1) and \ref{fig:evolution}d, e, f (2).}
    \label{fig:orbit_crf}
\end{figure}

Planetary mass and the size of the iron core are additional important factors influencing the amount and lifetime of volcanism. Planets with large cores tend to undergo less partial melting due to larger hydrostatic pressure gradients in the mantle, which prevent melt from reaching the surface \citep{noack2017VolcanismOutgassing}. Furthermore, the mantles of planets with large cores cool efficiently \citep{noack2017VolcanismOutgassing}, which reduces the time a planet is volcanically active (as already shown in Figs. \ref{fig:evolution}c and \ref{fig:evolution}d). This prevents a hothouse state in many cases and enables the existence of habitable planets at a wider range of orbital distances, mantle water contents and oxygen fugacities (Fig. \ref{fig:orbit_crf}). Due to the lower outgassing rates, however, in many cases the surface remains frozen. By contrast, planets with small cores show strong, long-lasting volcanic activity, which limits the potential to develop habitable conditions (Fig. \ref{fig:orbit_crf}). The same applies to planets with higher mass, where the higher mantle volume supports long-lived volcanism. For very thick mantles however, the viscosity of the deepest mantle can become so large that a non-convective, stagnant region is formed \citep{stamenkovic2012INFLUENCEPRESSUREDEPENDENT}, shrinking the active, convective part of the mantle (see e.g. the case of \SI{3}{\Me}, 30\% core shown in Fig. \ref{fig:mass_crf} in the Appendix). With respect to outgassing, this resembles the behavior of a smaller planet with a larger core.

\subsection{Planetary evolution pathways}
\begin{figure}[ht!]
    \centering
    \includegraphics[width=\linewidth]{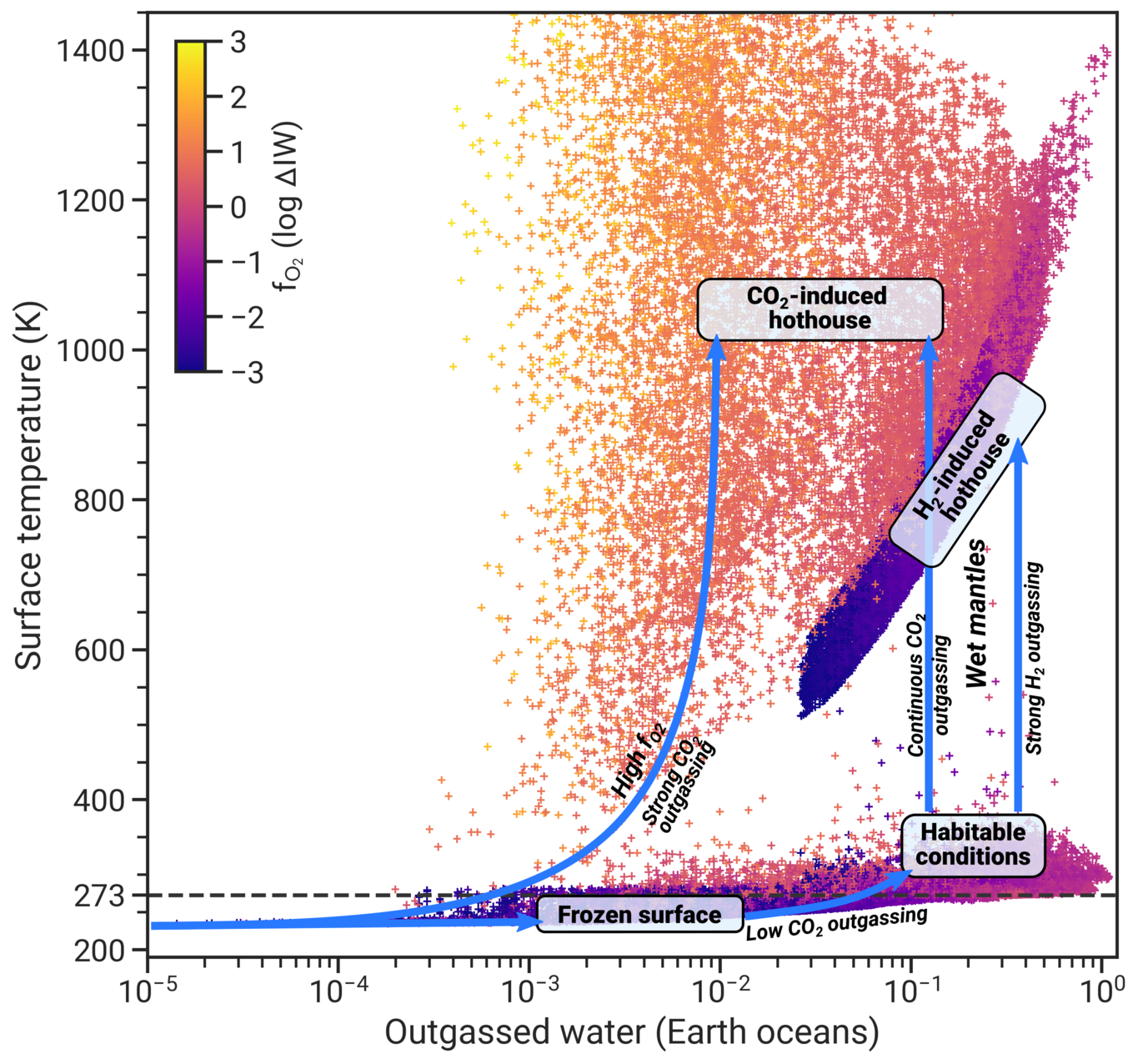}
    \caption{Evolutionary tracks of water outgassing on stagnant-lid planets. The arrows illustrate potential pathways on which a planet may evolve over time. Only planets with a non-zero amount of water outgassing are shown. The points represent Earth-mass planets with cores sizes ranging from 30\% to 70\% of the planet's radius at randomly selected times in their evolution, with the color showing the redox state of the mantle.}
    \label{fig:outgassedH2O_Ts}
\end{figure}

We find that distinct pathways exist for the evolution of a rocky, stagnant-lid planet (Fig. \ref{fig:outgassedH2O_Ts}), depending on the make-up of its mantle. If the planet's surface temperature remains low (e.g. for wide planetary orbits), outgassed water can start condensing and accumulating on the surface as ice or in liquid form. If \ce{CO2} outgassing is strong early on (e.g. for planets with oxidized mantles), this will quickly result in atmospheric pressures unsuitable for water outgassing. The planet develops a thick, hot, \ce{CO2}-rich atmosphere. By contrast, with weak \ce{CO2} outgassing, the planet can accumulate large amounts of water vapor, which will rapidly condense. Further \ce{CO2} outgassing can eventually push these planets into a hothouse regime, where the entire ocean evaporates to form a hot steam atmosphere. In planets with wet mantles and reducing conditions, outgassing of \ce{H2} can achieve the same effect. As discussed above, the weathering mechanisms on stagnant-lid planets do not permit a long-term removal of \ce{CO2}, preventing a return to habitable conditions even if all water vapor in the atmosphere were lost (a mechanism that we do not model here). Planets can stay in the habitable regime if the outgassing rates of greenhouse gases remain low over the volcanic lifetime of the planet, but high enough to keep the surface from freezing. This is the case for planets with oxygen fugacities around the IW buffer. Planets with dry, reduced mantles may never advance out of a frozen state due to limited outgassing of \ce{H2} and \ce{CO2}.

The presence of primordial atmospheres can reduce the range of habitable conditions even further. Substantial \ce{H2} atmospheres may not be lost quickly enough through atmospheric escape to allow the emergence of habitable surface conditions (Section \ref{sec:primordial_h2} in the Appendix). While pure steam atmospheres collapse quickly into an ocean, the presence of enough \ce{CO2} in a primordial atmosphere can strongly limit the range of interior conditions which yield habitable planets (Section \ref{sec:primordial_co2} in the Appendix).

\subsection{Sensitivity to model parameters}
\label{sec:sensitivity}

We tested the model sensitivity with respect to changing a few key parameters to confirm the robustness of the results. Fig. \ref{fig:sensitivity} summarizes this sensitivity analysis for planets with one Earth-mass.

\subsubsection{Influence of \ce{CO2} absorption coefficient}
\label{sec:sensitivity_co2}

Although our adopted value of the \ce{CO2} absorption coefficient $k_{0,\ce{CO2}}$ (\SI{0.05}{\m\squared\per\kg}) is chosen to reproduce the present-day Earth climate sensitivity \citep{pujol2003AnalyticalInvestigation}, which describes the response of Earth's climate to a doubling in \ce{CO2}, other values have been used in the literature \citep[e.g.,][]{elkins-tanton2008LinkedMagma}. Since the amount of greenhouse heating of \ce{CO2} plays an important role for habitability, we tested the sensitivity of the model to a reduction of $k_{0,\ce{CO2}}$ to \SI{0.001}{\m\squared\per\kg}, which is generally used in magma ocean atmosphere modeling \citep{nikolaou2019WhatFactors, elkins-tanton2008LinkedMagma} and thus provides us with a lower bound on the greenhouse heating from \ce{CO2}.
As shown in Fig. \ref{sec:sensitivity}b, this slightly extends the range of habitable planets towards more oxidizing conditions, as larger amounts of \ce{CO2} are needed for a planet to be in a hothouse regime due to the lower efficiency of heating. Overall, the influence of the absorption coefficient is fairly minor, which can largely be attributed to the self-regulating nature of the feedback mechanisms.

\subsubsection{Influence of ratio of intrusive/extrusive volcanism}
\label{sec:intrusive_volc}
\begin{figure}[h]
    \centering
    \includegraphics[width=\linewidth]{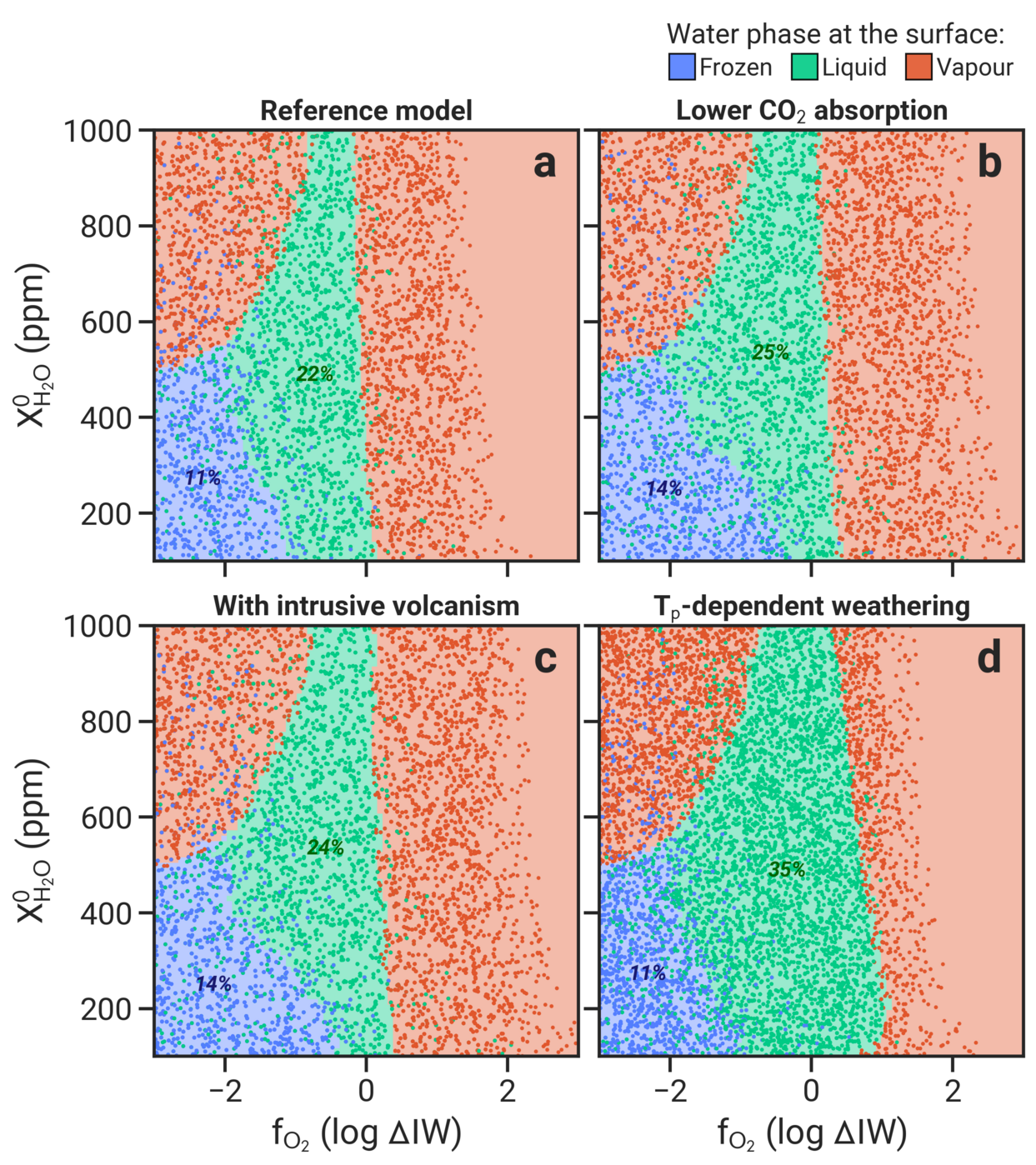}
    \caption{Influence of changing the \ce{CO2} absorption coefficient, introducing intrusive volcanism, and using temperature-dependent carbon weathering. The reference model shown here is the same as in Fig. \ref{fig:earth_4panels}.}
    \label{fig:sensitivity}
\end{figure}

In the models presented so far, we assumed that all melt produced in the mantle reaches the surface of the planet, where the supersaturated volatile species are outgassed into the atmosphere. However, in general a large part of the magma produced at depth is expected to be intrusive \citep{white2006LongtermVolumetric}, where melt crystallizes within or at the base of existing crust, thus reducing the amount of volatiles that reach the surface. The degree of extrusive volcanism ($f\_{extr}$) is difficult to constrain and can vary based on location, crust porosity and lithospheric thickness. To model the impact of reduced extrusive volcanism, we run a model study for an intrusive-to-extrusive ratio of 2.5 \citep{tosi2017HabitabilityStagnantlid}, corresponding to $f\_{extr}\approx0.286$. The results are shown in Fig. \ref{fig:sensitivity}c.

Similar to section \ref{sec:sensitivity_co2}, the presence of intrusive volcanism extends the range of habitable planets to more oxidizing conditions to a small degree. With intrusive volcanism, the rate of outgassing is reduced. This primarily reduces the rate at which greenhouse gases, specifically \ce{CO2}, build up.
Therefore, more \ce{CO2} can be outgassed until the planet's atmosphere is too hot too allow for liquid water, which allows the planets to be habitable at more oxidizing conditions.

\subsubsection{Sensitivity to carbon weathering efficiency}
The climate feedback from carbon weathering in the models presented here is fairly weak due to the sole dependence of the weathering model on the \ce{CO2} partial pressure, which we chose to provide a conservative estimate of the ermergence of habitable conditions. More efficient feedback mechanisms with a strong dependence on temperature have been suggested \citep[][]{krissansen-totton2017ConstrainingClimate}, which would be able to remove much more \ce{CO2} from the atmosphere and thus extend the range of habitable conditions.
To gauge the effect of temperature-dependent weathering, we replace the $\left(\frac{P_{\ce{CO2}}}{P_{\ce{CO2},\text{E}}}\right)^\alpha$ term in Eq. \eqref{eq:weathersl} with a temperature-dependent term $\exp{\left| \frac{E\_{bas}}{R} \left(\frac{1}{R T\_{p,E}} - \frac{1}{R T\_{p}}\right)\right]}$ following \citet{krissansen-totton2017ConstrainingClimate} and \citet{honing2019CarbonCycling}, where $E\_{bas}$ is the activation energy for basalt weathering, and $T\_{p}$ is the pore-space temperature which scales linearly with the surface temperature $T\_{s}$ as follows \citep{krissansen-totton2017ConstrainingClimate}:
\begin{equation}
    T\_p = a\_{grad} T\_s + b\_{int} + \SI{9}{\K},
\end{equation}
with constants $a\_{grad} = 1.02$ and $b\_{int} = \SI{-16.7}{\K}$. $T\_{p,E}$ is the pore-space temperature based on the current mean surface temperature $T\_{s,E}=\SI{285}{K}$ of Earth.

We find that the stronger feedback introduced by the temperature-dependence significantly increases the range of habitable planets towards more oxidizing mantle conditions (Fig. \ref{fig:sensitivity}d), in particular for mantles with low water contents. Compared to the model with \ce{CO2}-dependent weathering, the mantle water content plays a larger role. This is mainly due to the lower rate of volcanism on planets with drier mantles. The temperature-dependent weathering feedback works to establish a \ce{CO2} pressure in the atmosphere where the surface temperature is high enough so that the ingassed \ce{CO2} flux from weathering is in steady-state with outgassing fluxes from volcanism and decarbonation \citep[see also][]{honing2019CarbonCycling}. If the rate of volcanism is large (e.g. for high mantle water contents), the rates of outgassed and decarbonized \ce{CO2} are also large (since the latter is directly tied to weathering rates). In this case, a steady-state is not possible, as the required steady-state temperature would be too high for liquid water. Weathering therefore can not prevent a hothouse atmosphere. On the other hand, if the rate of volcanism is small, weathering can remove any excess \ce{CO2} at temperature ranges which permit liquid water on the surface. A comparison of the surface temperature and pressure evolution for both weathering models is shown in the appendix (Fig. \ref{fig:evolution_weathering}).

Overall, we find that the results and evolutionary paths shown above still hold well even with a stronger weathering feedback, although the exact range of  $f_{\ce{O2}}$ yielding habitable planets in Fig. \ref{fig:sensitivity} is different. As a particular side effect of the temperature-dependence of weathering, the \ce{CO2} content in the atmosphere is no longer driven solely by the rate of volcanism, but also by the increase in solar luminosity and other factors that increase the surface temperature of the planet, such as the water content of the atmosphere. We refer to \citet{honing2019CarbonCycling} for a more detailed discussion of the two weathering models for stagnant-lid planets.


\section{Discussion and conclusions}

The structure and composition of the interior are fundamental factors to determine whether a planet can be habitable or not. The mantle redox state in particular strongly constrains the space of potentially habitable planets. In general, it is difficult for planets with stagnant lids to remain habitable because of the limited number of pathways available to permanently remove outgassed \ce{CO2} from the atmosphere. In fact, \ce{CO2} tends to accumulate and heat the planet up until the atmosphere is too hot too allow liquid water. We model the (temporary) removal of \ce{CO2} via silicate weathering, which is active only in the presence of liquid water, thus placing further limits on the removal of \ce{CO2}. Additional \ce{CO2} could be lost through drag from escaping hydrogen \citep{tian2015HistoryWater, hunten1987MassFractionation} or through photodissociation in the upper atmosphere, processes we do not model here. Stagnant-lid planets are common in the Solar System. Earth alone is in a plate tectonic regime. If this is also the case for rocky exoplanets, we would expect a large number of those to more closely resemble Venus than Earth, with hot, dense atmospheres even if they reside in the habitable zone of their host star.

Our simple atmosphere model cannot capture the full complexity of a planetary atmosphere. With more sophisticated atmospheric models \citep{scheucher2020ConsistentlySimulating, wunderlich2020DistinguishingWet, kaspi2015AtmosphericDynamics, schreier2014GARLICGeneral}, the surface temperature would likely differ to some extent from the one calculated here, thus affecting both the habitability of planets and the critical amount of \ce{CO2} or \ce{H2} at which the planet transitions into a hot, non-habitable regime. However, as discussed above, the main driver for the accumulation of surface water is the outgassing rate of \ce{CO2} and \ce{H2}, which is mainly a function of the planetary interior. We note here that the outgassing rates are subject to the composition of the atmosphere, which may be different from the outgassed species due to atmospheric chemistry, which we do not take into account here. However, the solubilities of both \ce{CO2} and \ce{H2} in the melt are very low, and therefore these two species are least affected by partial pressures during outgassing. As such, while a difference in surface temperature could change the exact values of oxygen fugacities and water concentration in the mantle which yield habitable conditions, the general relations described above would still hold. As seen in Section \ref{sec:sensitivity_co2}, even a drastically reduced IR absorption coefficient of \ce{CO2} has only a small effect upon the proportions of planets with surface water.


Here we considered the internal structure of a planet to be independent of its redox state. Yet, the latter and the size of the metallic core may well evolve jointly, based on the local oxidation level of the protoplanetary disk during formation, on the conditions of metal-silicate differentiation, and on the subsequent evolution of the magma ocean, complex processes whose mutual relations are still to be fully unraveled \citep{wade2005CoreFormation, frost2008RedoxState, zhang2017EffectPressure, armstrong2019DeepMagma}. Planets with deep magma oceans may develop rather oxidizing mantles \citep{deng2020MagmaOcean}, which would be less favored to develop long-term habitable conditions based on our results. In fact, our results indicate that a stagnant-lid Earth or Venus, having more oxidizing conditions, will always enter a hothouse state. In contrast, low-mass planets with large iron cores would likely have more reducing conditions due to more shallow magma oceans, which in turn would strongly favor long-term habitable conditions. \citet{liggins2022GrowthEvolution} show that the mantle redox state imposes characteristic atmospheric compositions. The atmospheric composition of such planets is potentially detectable in exoplanets in the near-future via spectroscopic observations with instruments such as JWST \citep{greene2016CharacterizingTransiting}. Our results suggest that planets which are habitable should generally have reduced atmospheric compositions.
The unambiguous direct detection of an exoplanetary ocean is challenging, relying on direct imaging \citep{lustig-yaeger2018DetectingOcean}, and is therefore likely to be limited to only a few planets in the foreseeable future. Promising first targets would therefore be small, dense exoplanets, which may offer the best chances in the pursuit to find planets capable of hosting liquid water on their surface.

\begin{acknowledgements}
We thank Brad Foley for his thorough and constructive review that helped to significantly improve an earlier version of this manuscript. PB and NT acknowledge support of the German Science Foundation (DFG) through the priority program SPP 1992 "Exploring the Diversity of Extrasolar Planets" (TO 704/3-1) and the research unit FOR 2440 "Matter under planetary interior conditions" (PA  3689/1-1).
\end{acknowledgements}

\bibliographystyle{bibtex/aa} 
\bibliography{manuscript.bib}

                                    
\begin{appendix}

\section{1D parameterized convection model}
\label{sec:convection_model}

\begin{figure}[ht!]
    \centering
    \includegraphics[width=\linewidth]{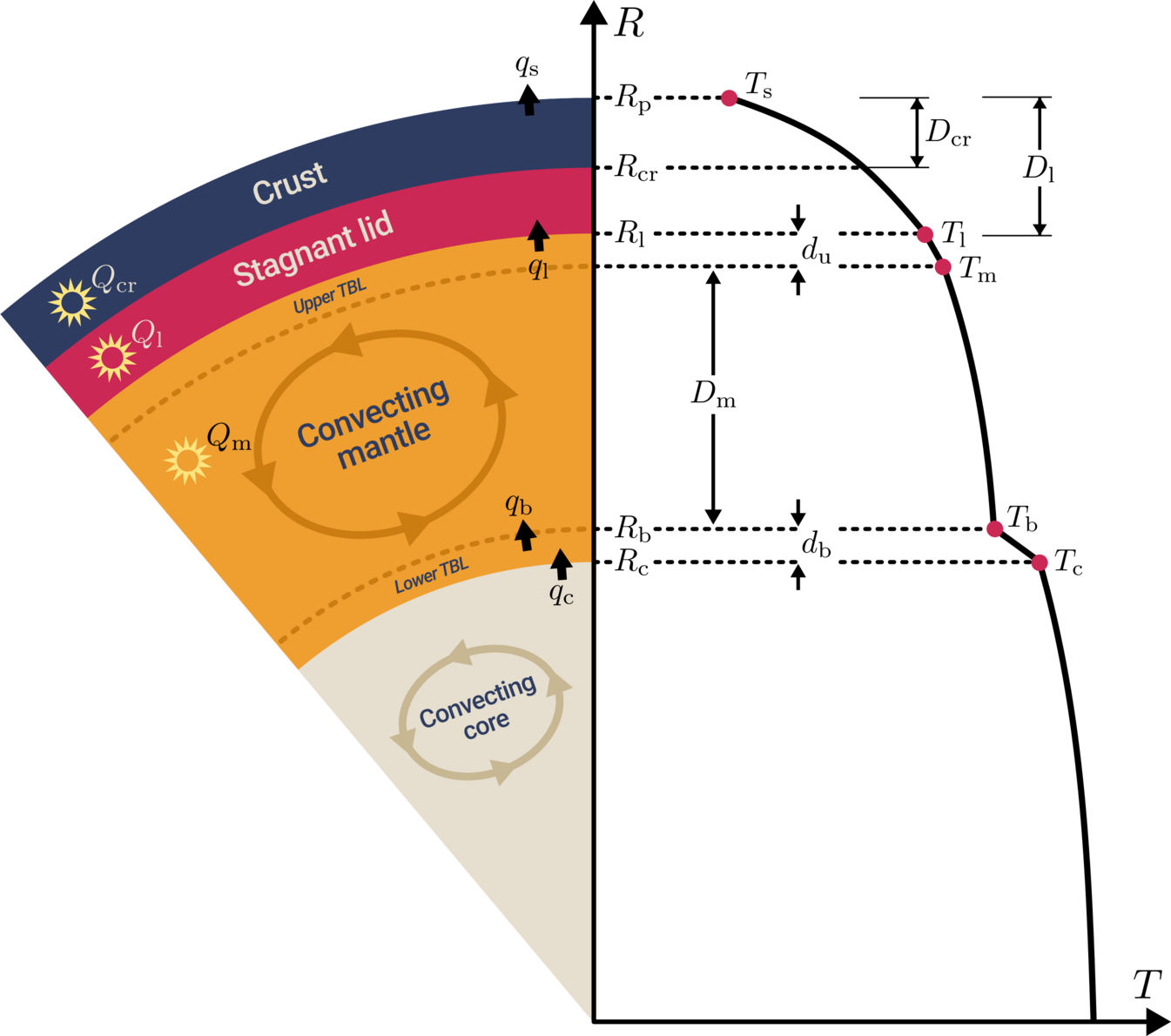}
    \caption{Schematic of the interior structure used in the thermal evolution model, alongside a diagram of the temperature profile (see the text for the explanation of the various symbols).}
    \label{fig:model_structure}
\end{figure}

We model the thermal evolution of a planet's mantle by considering the energy balance between heat lost through the planetary surface and heat entering the mantle from the iron core, as well as the decay of radiogenic elements inside the mantle. Assuming the core to be fully liquid and convecting, its energy balance is given by
\begin{equation}
    \label{eq:dTc}
    \rho\_c c\_c V\_c  \dt{\overbar{T\_c}} = - q\_c A\_c,
\end{equation}
where $\overbar{T\_c}$ is the average temperature in the core; $\rho\_c$, $c\_c$, and $V\_c$ are the density, specific heat capacity, and volume of the core; $A\_c$ is the area of the core-mantle boundary (CMB); and $q\_c$ is the heat flux out of the core.

The energy balance of the mantle is given by
\begin{equation}
    \label{eq:dTm}
    \begin{split}
    &\rho\_m c\_m V\_m (1+\St) \dt{\overbar{T\_m}} =\\& - \left(q\_l + \left[\rho_{cr} \mathcal{L} + \rho\_{cr} c\_{cr} (T\_m - T\_l) \right] \dt{D\_{cr}} \right) A\_m + q\_b A\_b + Q\_m V\_m,
    \end{split}
\end{equation}
where $\overbar{T\_m}$ is the average temperature of the convecting mantle; $\rho\_m$ and $c\_m$ are the density and specific heat capacity of the mantle; $V\_m$ and $A\_m$ are the volume and area of the convecting part of the mantle; $\St$ is the Stefan number, which describes the energy consumed and released upon mantle melting and solidification; $q\_l$ and $q\_b$ are the heat fluxes out of and into the convecting part of the mantle respectively; $T\_m$ and $T\_l$ are the temperatures of the upper mantle and the bottom of the stagnant lid; $\rho_{cr}$, $c\_{cr}$, and $D\_{cr}$ are the density, specific heat capacity, and thickness of the crust; $\mathcal{L}$ is the latent heat of melting; and $Q\_m$ is the volumetric heating rate in the mantle from heat-producing elements. The main parameters used in this model are summarized in Table \ref{tab:parameters}.

The temperatures at the top of the core and mantle $T\_c$ and $T\_m$ are related to the volume-averaged temperatures $\overbar{T\_c}$ and $\overbar{T\_m}$ through scaling factors $\varepsilon\_c$ and $\varepsilon\_m$
\begin{equation}
    \label{eq:avgTmTc}
        \overbar{T\_c} = \varepsilon\_c T\_c, \qquad \overbar{T\_m} = \varepsilon\_m T\_m = \frac{1}{V\_m} \Int{V\_m}{}{T\_{ad}(r)}{V},
\end{equation}
where $T\_{ad}(r)$ is the adiabatic temperature profile in the mantle. For the core, we set $\varepsilon\_c = 1.2$ following \citet{stamenkovic2012INFLUENCEPRESSUREDEPENDENT}, who found only a small dependence on planetary mass. For the mantle, $\varepsilon\_m$ is updated continuously during the mantle evolution based on the mantle temperature profile and on the varying thickness of the stagnant lid.

The evolution of the thickness of the stagnant lid ($D\_l$) follows from the energy balance at the base of the lid (at radius $R\_l$) 
\begin{equation}
    \label{eq:dDl}
    \begin{split}
        &\rho\_m c\_m V\_m (T\_m - T\_l) \dt{D\_l} =\\& -q\_l + \left[\rho_{cr} \mathcal{L} + \rho\_{cr} c\_{cr} (T\_m - T\_l) \right] \dt{D\_{cr}} - k\_m  \del{T}{r} \biggr\rvert_{r = R\_l}.
    \end{split}
\end{equation}
$\partial T / \partial r \rvert_{r = R\_l}$ is the temperature gradient at the base of the lid, which we calculate assuming steady-state heat conduction:
\begin{equation}
    \label{eq:conduction_lid}
    \frac{1}{r^2} \del{}{r} \left(r^2 k\_l \del{T}{r}\right) = - Q\_l,
\end{equation}
where $k\_l$ is the thermal conductivity and $Q\_l$ the heat production rate in the lid. Parts of the stagnant lid can be comprised of crustal material, so when solving Eq. \eqref{eq:conduction_lid}, we set $k\_l$ and $Q\_l$ to crustal values ($k\_{cr}$ and $Q\_{cr}$) from the surface to the base of the crust or mantle values ($k\_m$ and $Q\_m$) from the base of the crust to the base of the stagnant lid as appropriate. The boundary conditions for Eq. \ref{eq:conduction_lid} are the given surface temperature $T\_s$ and the temperature at the stagnant lid base $T\_l$.

Numerical convection models suggest that the viscosity contrast across the upper thermal boundary layer is typically about one order of magnitude \citep{grasset1998ThermalConvection}. Based on this, the lid base temperature $T\_l$ can be calculated from the mantle temperature and activation energy \citep{grasset1998ThermalConvection}

\begin{equation}
    \label{eq:Tl}
    T\_l = T\_m - 2.9 \frac{R T^2\_m}{E^*},
\end{equation}
where the factor of 2.9 accounts for the effects of spherical geometry \citep{reese2005ScalingLaws}.

The convective heat flux out of the mantle $q\_l$, assuming that the upper thermal boundary layer is small so that the radial temperature profile is close to linear, can then be expressed as
\begin{equation}
    \label{eq:ql}
    q\_l = k\_m \frac{T\_m - T\_l}{d\_u},
\end{equation}
where the thickness of the upper TBL $d\_u$ can be calculated from boundary layer theory
\begin{equation}
    \label{eq:du}
    d\_u = D\_m \left( \frac{\Ra\_{crit}}{\Ra}\right)^{\nicefrac{1}{3}},
\end{equation}
where $D\_m=R\_l - R\_b$ is the thickness of the convecting part of the mantle, $\Ra\_{crit}$ is the critical Rayleigh number of the mantle, and $\Ra$ is the Rayleigh number for the entire mantle
\begin{equation}
    \label{eq:Ra}
    \Ra = \frac{\alpha\_m(P\_m, T\_m) \rho\_m g \Delta T D\_m^3}{\kappa\_m \eta\_m},
\end{equation}
with the mantle viscosity $\eta\_m$, mantle thermal diffusivity $\kappa\_m = k\_m / (\rho\_m c\_m)$, the pressure- and temperature-dependent coefficient of thermal expansion $\alpha\_m$ with the pressure at the top of the mantle $P\_m=\rho\_m g (D\_l + d\_u)$, and $\Delta T = (T\_m - T\_l) + (T\_c - T\_b)$, which is the sum of temperature differences across both boundary layers.

The coefficient of thermal expansion $\alpha$, which has a strong influence on the heat transport in the convecting mantle, is often assumed to be constant, although it is known from experimental data that this parameter can vary considerably with both pressure and temperature \citep{fei1995ThermalExpansion}. This becomes especially important if one considers the modeling of super-Earths \citep{miyagoshi2018EffectsAdiabatic}. We use here the temperature- and pressure-dependent parameterization of $\alpha$ from \citet{tosi2013MantleDynamics},
\begin{equation}
    \label{eq:alpha}
    \alpha (P, T) = (a_0 + a_1 T - a_2 T^{-2}) \exp(- a_3 P),
\end{equation}
where the coefficients $a_0$ - $a_3$ are chosen assuming a lower mantle composition of 80\% perovskite/20\% periclase (see Table \ref{tab:parameters} for parameter values).

The temperature profile in the convecting part of the mantle is assumed to be adiabatic:
\begin{equation}
    \label{eq:adiabat}
    \frac{\mathrm{d} T}{\mathrm{d} P} = \frac{\alpha(P, T)}{\rho\_m c\_m}\: T.
\end{equation}

To account for a potentially non-convecting zone near the CMB due to the effect of high pressures on the mantle viscosity, we use the parameterization from \citet{stamenkovic2012INFLUENCEPRESSUREDEPENDENT}. Assuming that this conductive layer is close to convective stability, the thickness can be approximated from boundary layer theory using a critical Rayleigh number $\Ra^{\text{CMB}}\_{crit}$ based on the viscosity contrast $\Delta\eta$ across the layer:

\begin{equation}
    \label{eq:Ra_cr}
    \Ra^{\text{CMB}}\_{crit}(\Delta\eta) = \max \bigl\{ \Ra\_{crit}, 11.74 \cdot \ln(\Delta\eta)^4 \bigr\},
\end{equation}
with
\begin{equation}
    \label{eq:viscosity_contrast}
    \Delta\eta = \max \left\{\frac{\eta(R\_c)}{\eta(R\_b)}, \frac{\eta(R\_b)}{\eta(R\_c)}\right\}.
\end{equation}
Here, $R\_b$ is the radius at the top of the conductive layer, and $R\_c$ is the radius of the CMB.

The thickness of this layer is then given by:
\begin{equation}
    \label{eq:dc}
    d\_b = \left( \Ra^{\text{CMB}}\_{crit}(\Delta\eta)\frac{\kappa\_m \min \bigl\{ \eta(R\_b), \eta(R\_c) \bigr\}}{\alpha\_m(P\_b, T\_b) \rho\_m g |T\_c - T\_b|}\right)^{\nicefrac{1}{3}}.
\end{equation}

Especially for planets more massive than Earth, the conductive CMB layer can make up a significant part of the mantle. Therefore, we treat the heat fluxes from the CMB lid into the mantle and from the core into the CMB lid separately, and assume time-dependent thermal conduction across the layer. The heat fluxes are given by the temperature gradients at the top and bottom of the conductive CMB layer:
\begin{equation}
    \label{eq:qb}
    q\_b = -k\_m \del{T}{r} \biggr\rvert_{r = R\_b}
\end{equation}

\begin{equation}
    \label{eq:qc}
    q\_c = -k\_m \del{T}{r} \biggr\rvert_{r = R\_c},
\end{equation}

which we determine by solving the time-dependent heat equation across the CMB lid:
\begin{equation}
    \label{eq:conduction_cmb}
    \frac{1}{r^2} \del{}{r} \left(r^2 k\_m \del{T}{r}\right) = - Q\_m + \rho\_m c\_m \del{T}{t}.
\end{equation}

\section{Melting, trace element partitioning, and volatile outgassing}
\label{sec:melting}

We compute the distribution of partial melt in the mantle by comparing the local mantle temperature profile $T(r)$ against the solidus $T\_{sol}(r)$ and liquidus $T\_{liq}(r)$ temperatures. We assume the amount of partial melt to vary linearly between the solidus and the liquidus:
\begin{equation}
    \label{eq:mf}
    \phi(r) = \frac{T(r) - T\_{sol}(r)}{T\_{liq}(r) - T\_{sol}(r)}.
\end{equation}

We do not consider melting above a pressure of \SI{8}{GPa}. Under these conditions, melt becomes denser than the surrounding mantle rocks and cannot reach the surface \citep{agee2008StaticCompression}.

The presence of water in the mantle depresses the solidus and liquidus curves. We calculate wet solidus and liquidus curves following a parameterization by \citet{katz2003NewParameterization}.

The volume-averaged, extractable melt fraction $\overline{\phi}$ in the mantle is then given by
\begin{equation}
    \label{eq:avg_mf}
    \overline{\phi} = \frac{1}{V_\phi} \Int{V_\phi}{}{\phi(r)}{V},
\end{equation}
where $V_\phi$ is the total volume of the melt zone (i.e. where the temperature lies above the solidus).

Knowing the volume of melt produced, we can calculate the evolution of the crustal thickness $D\_{cr}$. We adopt the plume model description by \citet{grott2011VolcanicOutgassing}. Partial melting in the mantle is generally restricted to localized upwelling plumes. We assume a plume covering fraction of $f\_p = 0.01$, and add the temperature difference across the bottom thermal boundary layer to the local temperature profile when evaluating the melt fraction in Eq. \eqref{eq:mf}. In addition to the amount of available melt, the crustal growth rate depends on the rate at which fresh mantle material can be supplied to the partial melt zone, which is governed by the convective velocity $u$ of the mantle. The crustal growth rate is given by
\begin{equation}
    \label{eq:dDcr}
    \dt{D\_{cr}} = f\_p u \overline{\phi} \frac{V_\phi}{4 \pi R\_p^3},
\end{equation}

where the convective velocity is
\begin{equation}
    \label{eq:conv_vel}
    u = u_0 \left( \frac{\Ra}{\Ra\_{crit}}\right)^{\nicefrac{2}{3}},
\end{equation}
where $u_0$ is a characteristic mantle convective velocity scale.
We impose the additional constraint that the crust cannot grow larger than the lid. Once the crust reaches the bottom of the lid, any excess crust is recycled back into the mantle, and the crustal growth rate is set to be equal to the lid growth rate.

During crustal formation, we treat the release and consumption of latent heat during mantle melting and crystallization via the Stefan number, which is recalculated at every time step (See also Eq. \eqref{eq:dTm})
\begin{equation}
    \label{eq:St}
    \St = \frac{\mathcal{L}}{c\_m} \frac{V_\phi}{V\_m} \frac{\mathrm{d} \overbar{\phi}}{\mathrm{d} T\_m}.
\end{equation}

We consider the partitioning of radiogenic elements and water between crust and mantle due to melt production and crust formation, and the subsequent enrichment of the crust in these elements. We consider a model of fractional melting to calculate the partitioning of trace elements between melt and mantle rocks. The concentration in the melt $X\_{liq}^i$ of a given trace element $i$ at radius $r$ is then given by
\begin{equation}
    \label{eq:X_liq}
    X\_{liq}^i(r) = \frac{X\_m^i}{\phi(r)}\left[1 - \big(1-\phi(r)\big)^{\nicefrac{1}{\delta_i}}\right],
\end{equation}
where $X\_m^i$ is the corresponding bulk concentration in the mantle and $\delta_i$ a trace-element-specific partition coefficient. We assume $\delta_i=0.001$ for heat-producing elements \citep{blundy2003PartitioningTrace}, and $\delta_i=0.01$ for water \citep{aubaud2004HydrogenPartition}. The average concentration in the melt then follows as

\begin{equation}
    \label{eq:X_liq_avg}
    \overbar{X}\_{liq}^i = \frac{1}{\overline{\phi} V_\phi} \Int{V_\phi}{}{X\_{liq}^i(r) \phi(r)}{V}.
\end{equation}

Enriched melt is transported to the surface and forms a crust. The total mass of an incompatible element $M\_{cr}^i$ in the crust is given by the crust production rate and the average concentration in the melt:
\begin{equation}
    \label{eq:dMcr_i}
    \dt{M\_{cr}^i} = 4 \pi R\_p^2 \rho\_{cr} \overbar{X}\_{liq}^i \dt{D\_{cr}}
\end{equation}
At the same time, the mantle will be depleted in trace elements accordingly, with the concentration of the trace elements in the mantle and crust given by
\begin{equation}
    \label{eq:X_m_i}
    X\_m^i = \frac{X\_{m,0}^i M\_{m,0} - M\_{cr}^i}{M\_m}, \quad X\_{cr}^i = \frac{M\_{cr}^i}{M\_{cr}},
\end{equation}
where $M\_{m,0}$ and $M\_m$ are the initial and current mass of the mantle, respectively, and $X\_{m,0}^i$ is the initial mantle concentration of the respective trace element.

The enrichment of heat-producing elements in the crust and depletion in the mantle leads to different volumetric heating rates in crust and mantle, which can be calculated as follows
\begin{equation}
    \label{eq:Qm}
    Q\_m(t) = \rho\_m \sum_i X\_m^i(t) H_i \exp\left(\frac{\ln 2 \cdot (\SI{4.5}{Gyr} - t)}{\tau_{\nicefrac{1}{2}}^i}\right),
\end{equation}

\begin{equation}
    \label{eq:Qcr}
    Q\_{cr}(t) = \rho\_{cr} \sum_i X\_{cr}^i(t) H_i \exp\left(\frac{\ln 2 \cdot (\SI{4.5}{Gyr} - t)}{\tau_{\nicefrac{1}{2}}^i}\right),
\end{equation}
where $i$ specifies one of the four long-lived radioisotopes \ce{^40 K}, \ce{^232 Th}, \ce{^235 U} and \ce{^238 U}, with corresponding specific heat production rates $H_i$ and half-lives $\tau_{\nicefrac{1}{2}}^i$ based on bulk-silicate-Earth abundances from \citet{mcdonough1995CompositionEarth}.



Not all melt produced in the mantle is able to reach the surface, but is instead intruded into the lid, solidifies there, and is therefore unavailable for outgassing. To model this, we need to assume a fraction of extrusive volcanism $f\_{extr}$, which we set to 1 in this study (i.e. all melt reaches the surface). While this is not fully realistic for an Earth-like planet, it provides an upper limit for the outgassed species. In Sect. \ref{sec:intrusive_volc} in the main text, we also tested the model results with a more realistic value \citep{tosi2017HabitabilityStagnantlid} of $f\_{extr}=0.286$ and show that this ultimately serves to further increase the number of habitable planets.

Volatile species will be partially outgassed into the atmosphere once the melt reaches the surface. To outgas a volatile species, the melt needs to be supersaturated with respect to the atmosphere, and any excess concentration can be released into the atmosphere. The redox state of the mantle plays a large role regarding which volatile species are outgassed, with an oxidized mantle favoring \ce{H2O} and \ce{CO2}, while a reduced mantle is dominated by \ce{H2} and \ce{CO} outgassing \citep{ortenzi2020MantleRedox}. The chemical outgassing model based on \citet{ortenzi2020MantleRedox} (as described in Section \ref{sec:outgassing_model}) calculates the outgassed mass fraction $X\_{outg}^i$ of volatile species based on the chemical equilibrium between melt and atmosphere, taking into account the solubility of each species.
The outgassed mass $M\_{outg}^i$ of volatile species is then given by

\begin{equation}
    \label{eq:dM_outg}
    \dt{M\_{outg}^i} = f\_{extr} X\_{outg}^i \dt{M\_{cr}^i}.
\end{equation}

We can then calculate the partial pressure of each species according to its atmospheric mass and the presence of other species \citep[e.g.][]{bower2019LinkingEvolution}. The actual mass enriched in the crust in Eq. \eqref{eq:dMcr_i} is then reduced by the outgassed mass. We also assume that all surface volcanism takes place above any potential ocean surface, since the pressure at the bottom of the ocean would limit outgassing. Ocean coverage and depth are difficult to estimate as they are dependent on surface topography. Similar to the assumption of fully extrusive volcanism, this assumption provides an upper limit to volatile outgassing. Likewise, we do not take into account so-called \enquote{water worlds}, i.e. planets with several tens of kilometers of water oceans. Due to the high pressures at the ocean bottom, volatile outgassing would likely be severely limited \citep{noack2016WaterrichPlanets, krissansen-totton2021WaterworldsProbably}.

\section{Venus-like planets}
\label{sec:venus}
\begin{figure}[ht!]
    \centering
    \includegraphics[width=\linewidth]{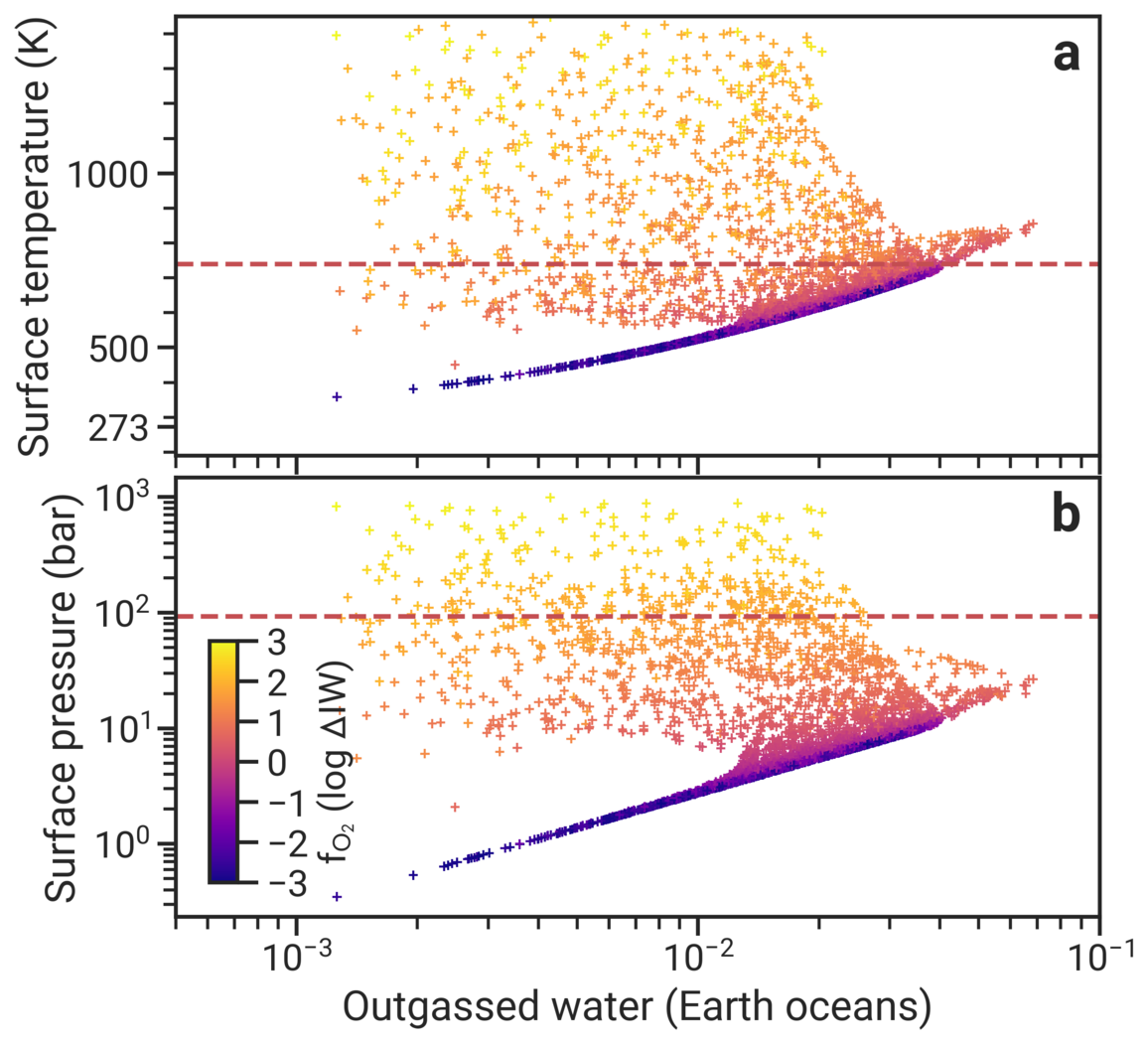}
    \caption{Thermal (a) and pressure (b) state of Venus-like planets at present day, after \SI{4.5}{\giga\year} of evolution. The color map indicates the oxygen fugacity of the mantle. The red dashed lines mark the present-day surface temperature and atmospheric pressure of Venus.}
    \label{fig:venus_currentday_TP}
\end{figure}

Venus is the quintessential hothouse planet. In order to test the ability of our model to reproduce a Venus-like scenario, we simulated the evolution of 5000 planets with Venus-like interior structures and orbital distance, while varying the mantle water content and oxygen fugacity as described in the methods section \ref{sec:initial_parameters}. We find that at present day (\SI{4.5}{\giga\year}) all modeled planets are in an extreme greenhouse state. No habitable planets are present (Fig. \ref{fig:venus_currentday_TP}a), and surface pressures in planets with mantle oxygen fugacities above the IW buffer are comparable to those of present-day Venus (Fig. \ref{fig:venus_currentday_TP}b). Our model tends to overestimate the surface temperature compared to actual Venus. This stems from the fact that we do not consider the loss of water through photodissociation, which leaves water in the atmosphere as a potent greenhouse gas. Many of the water-rich atmospheres shown in Fig. \ref{fig:venus_currentday_TP} would evolve into thick, dry, \ce{CO2}-dominated atmospheres.
Furthermore, Venus' high albedo due to its global cloud cover is not represented in the model, which contributes to explaining the higher surface temperatures. A more in-depth discussion of the evolution of Venus using the outgassing model discussed here can be found in \citet{honing2021EarlyHabitability}. 

\section{Influence of primordial \ce{H2} atmospheres}
\label{sec:primordial_h2}

\begin{figure*}[htb!]
    \centering
    \includegraphics[width=0.9\linewidth]{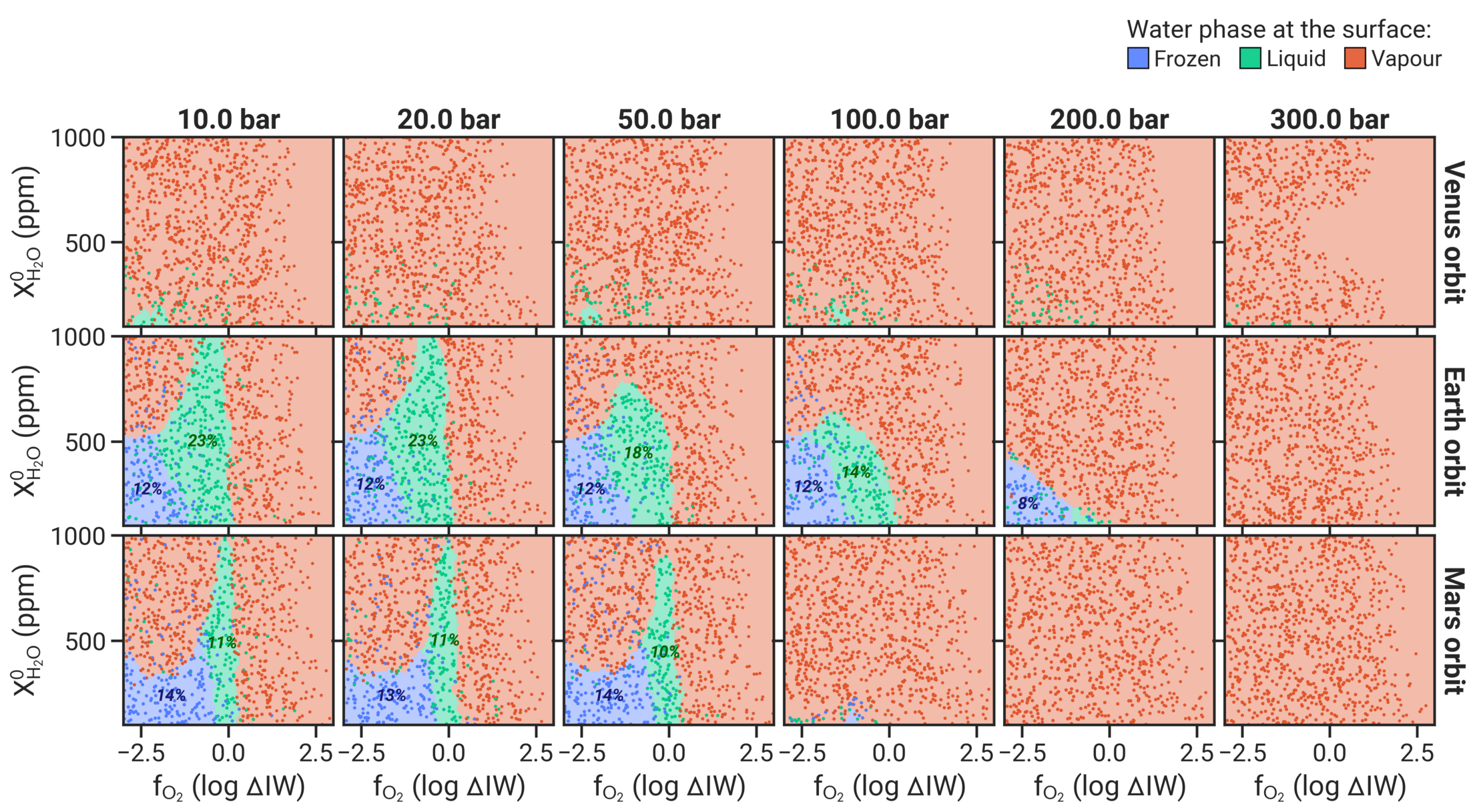}
    \caption{Influence of a primordial \ce{H2} atmosphere on the emergence of habitable surface conditions. Each column corresponds to different initial pressures of \ce{H2}, with each row marking planets at different orbital distances to their host star.}
    \label{fig:primordial_h2}
\end{figure*}

So far we have assumed that any primordial atmosphere is lost at the point when our evolution models start. This provides us with an upper limit to any outgassed secondary atmosphere. However, while the life-time of primordial atmospheres can be very short, on the order of tens of millions of years, especially for close-in, low-mass planets \citep{kite2020ExoplanetSecondary, lammer2014OriginLoss}, thick \ce{H2} atmospheres may survive magma ocean solidification. We investigated the influence of a primordial \ce{H2} atmosphere by running a number of evolution models of Earth-like planets with initial atmospheric pressures of up to \SI{300}{bar}. This upper limit is motivated by the amount of hydrogen an Earth-like planet may accrete on formation \citep{lammer2014OriginLoss} and by the maximum amount it can lose over time so that most planets we consider here are still rocky \citep{howe2020SurvivalPrimordial}. This results in a reasonable range of different planet evolutions, but different choices could change the proportions of planets with different atmospheres. Low-mass planets in particular may not be able to accrete a hydrogen atmosphere of that extent during formation.
We find that the presence of primordial \ce{H2} atmospheres with pressures above $\approx 50-\SI{100}{bar}$ can significantly reduce the amount of habitable planets (Fig. \ref{fig:primordial_h2}). This is due to two main factors: First, sufficiently thick \ce{H2} atmospheres may not be lost completely, and thus the planet never reaches surface temperatures that are low enough to allow for liquid water. Second, it can easily take several hundred million years for extensive primordial \ce{H2} atmospheres to be completely lost. At this early stage, a planet is volcanically very active, but the outgassed \ce{CO2} is not removed from the atmosphere because the existing \ce{H2} atmosphere inhibits the carbon-silicate cycle that requires liquid water. In addition, additional outgassing of \ce{H2} can further prolong the lifetime of an \ce{H2} atmosphere. As a result, even though the \ce{H2} atmosphere may be eventually lost, too much \ce{CO2} has been outgassed during that time to allow habitable conditions. The planets which may avoid a hothouse in these cases are those with very low amounts of \ce{CO2} outgassing, i.e. planets with dry mantles and oxygen fugacities well below the IW buffer.

\section{Influence of primordial steam and \ce{CO2} atmospheres}
\label{sec:primordial_co2}
\begin{figure*}[ht!]
    \centering
    \includegraphics[width=0.9\linewidth]{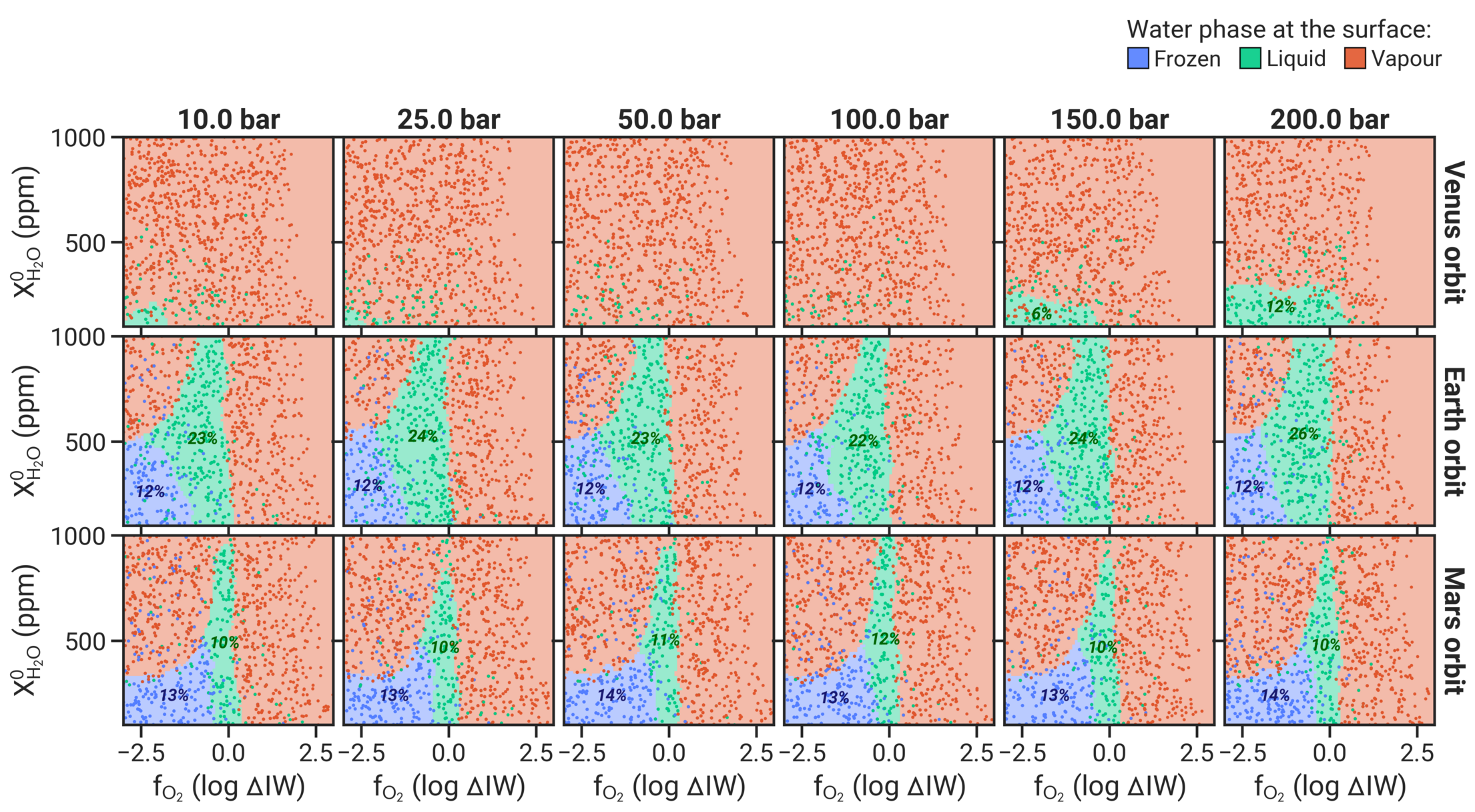}
    \caption{Influence of a primordial steam atmosphere on the emergence of habitable surface conditions. Each column of plots corresponds to different initial pressures of \ce{H2O}, with each row marking planets at different orbital distances to their host star.}
    \label{fig:primordial_h2o}
\end{figure*}

\begin{figure*}[ht!]
    \centering
    \includegraphics[width=0.9\linewidth]{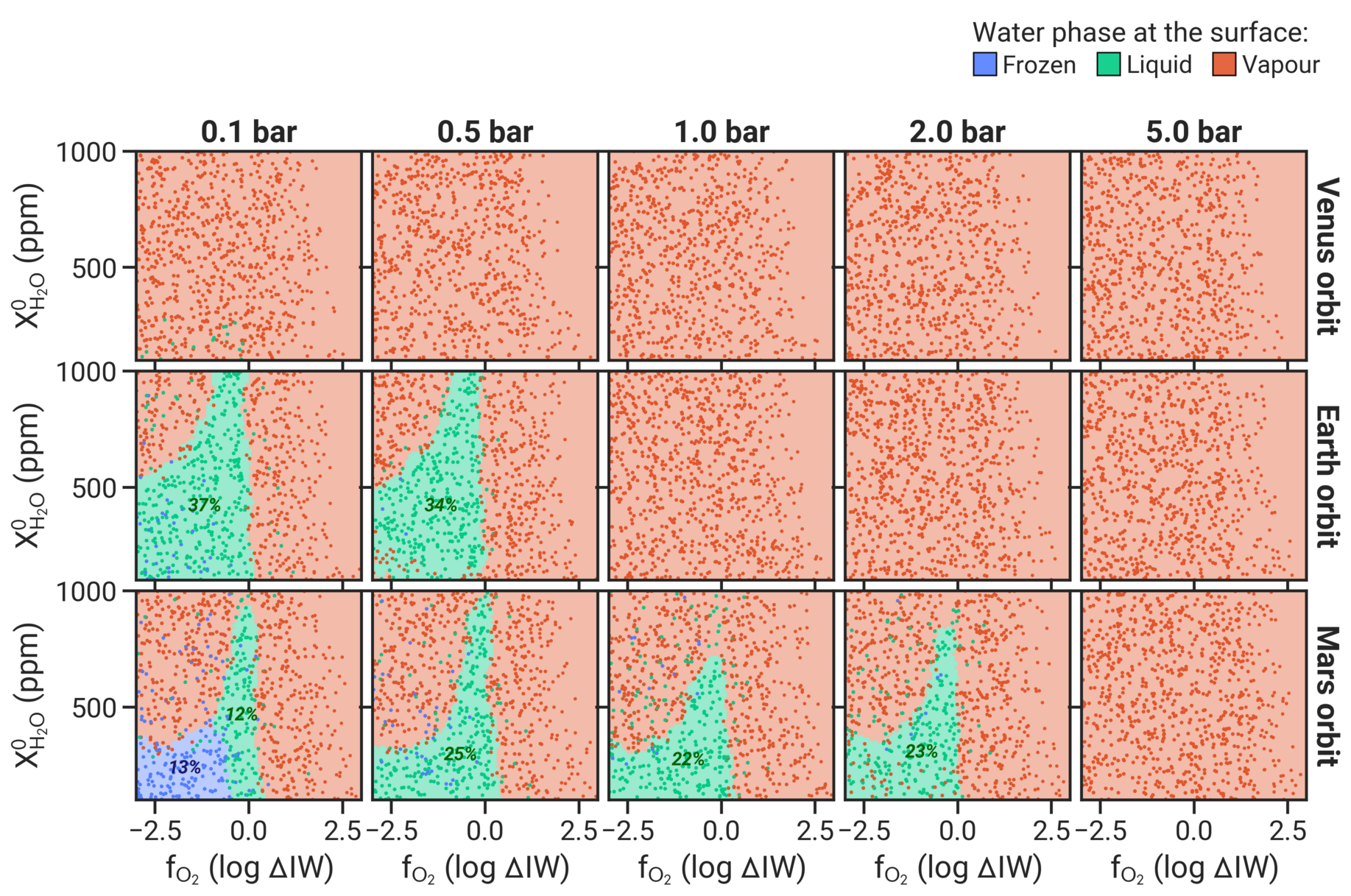}
    \caption{Influence of a primordial \ce{CO2} atmosphere on the emergence of habitable surface conditions. Each column of plots corresponds to different initial pressures of \ce{CO2}, with each row marking planets at different orbital distances to their host star.}
    \label{fig:primordial_co2}
\end{figure*}

It is likely that a magma ocean would form a thick steam atmosphere, although these may collapse shortly after magma ocean solidification to form an early ocean \citep{elkins-tanton2011FormationEarly}. To determine the influence of an early post magma ocean atmosphere, we investigate two end member compositions: Pure steam atmospheres up to \SI{200}{bar}, and pure \ce{CO2} atmosphere up to \SI{5}{bars}.
Pure steam atmospheres have little impact on the long-term evolution of the planets (Fig. \ref{fig:primordial_h2o}). Due to the positive climate feedback of water vapor, these atmospheres are not stable, and other sources of heating, such as from other greenhouse gases, are needed to sustain a steam atmosphere. Shortly after the start of the evolution, these atmospheres therefore collapse and rain out to form an ocean. They do not contribute further to the warming of the surface, but merely provide a reservoir of water.
On the other hand, even small amounts of initial \ce{CO2} atmospheres can severely limit the occurrence of habitable conditions (Fig. \ref{fig:primordial_co2}). In our model, \ce{CO2} is only removed via silicate weathering, which requires the presence of liquid water. If the initial amount of \ce{CO2} in the atmosphere is too high to permit liquid water, there exists no pathway for a planet to lose \ce{CO2} and become habitable.  Therefore, the \ce{CO2} pressures given in Fig. \ref{fig:primordial_co2} represent a conservative estimate on the amount of \ce{CO2} which can still yield habitable conditions.

\section{Tables}

\renewcommand{\arraystretch}{1}
\begin{table}[ht!]
\centering
\caption{Magma composition for the outgassing speciation model}
\begin{tabular}{lS[table-format=3.2]}
\hline
Component & {wt\%}\\
\hline
\ce{SiO2} & 49.9 \\
\ce{Al2O3} & 15.9\\
\ce{FeO} & 11.1\\
\ce{MgO} & 6.8\\
\ce{CaO} & 9.6\\
\ce{Na2O} & 3.0\\
\ce{TiO2} & 1.9 \\
\ce{K2O} &  1.2\\
\hline
\end{tabular}
\label{tab:magma_composition}
\end{table}

\clearpage
\newpage
\onecolumn
\renewcommand{\arraystretch}{1}
\newcolumntype{L}{>{\raggedright\arraybackslash\setlength{\baselineskip}{0.6\baselineskip}}X}%
\begin{center}
\begin{tabularx}{\textwidth}{cLlL}
\caption{Main model parameters}\label{tab:parameters}\\
\toprule
Parameter    & Description         & Value & Reference \\
\toprule \endhead
\midrule
\multicolumn{4}{r}{\itshape continues on next page}\\
\midrule\endfoot

\bottomrule\endlastfoot
\multicolumn{4}{l}{\textit{Mantle convection}} \\[0.2em]
$\rho\_{cr}$ & Crust density       & \SI{2900}{\kg\per\m\cubed}&\\
$k\_{cr}$    & Crust thermal conductivity & \SI{3}{\W\per\m\per\K}&\\
$k\_{m}$    & Mantle thermal conductivity & \SI{4}{\W\per\m\per\K}&\\
$c\_{cr}$    & Crust specific heat capacity & \SI{1100}{\J\per\kg\per\K}&\\
$c\_{m}$    & Mantle specific heat capacity & \SI{1100}{\J\per\kg\per\K}&\\ 
$c\_{c}$    & Core specific heat capacity & \SI{800}{\J\per\kg\per\K}&\\ 
$\Ra\_{crit}$  & Critical mantle Rayleigh number & 450&\\ 
$A$  & Viscosity pre-factor & \SI{6.127e10}{\Pa\s}& \citet{tosi2017HabitabilityStagnantlid}\\
$E^*$  & Activation energy & \SI{3.35e5}{\J\per\mol}& \citet{tosi2017HabitabilityStagnantlid}\\
$a_0$  & Thermal expansivity coefficient & \SI{2.68e-5}{\per\K}& \citet{tosi2013MantleDynamics}\\
$a_1$  & Thermal expansivity coefficient & \SI{2.77e-9}{\per\K\squared}& \citet{tosi2013MantleDynamics}\\
$a_2$  & Thermal expansivity coefficient & \SI{-1.21}{\K}&  \citet{tosi2013MantleDynamics}\\
$a_3$  & Thermal expansivity coefficient & \SI{8.63e-3}{\per\giga\Pa}& \citet{tosi2013MantleDynamics}\\
\midrule
\multicolumn{4}{l}{\textit{Melting}} \\[0.2em]
$u_0$  & Convection velocity scale & \SI{2e-12}{\m\per\s}& \citet{spohn1991MantleDifferentiation}\\ 
$\mathcal{L}$  & Latent heat of melting & \SI{6e5}{\J\per\kg}&\\
$f\_p$  & Plume covering fraction & 0.01& \citet{tosi2017HabitabilityStagnantlid}\\
$f\_{extr}$  & Proportion of extrusive volcanism & 1.0 or 0.286&\\
$\delta_{\ce{H2O}}$ & Water partition coefficient & 0.01 & \citet{aubaud2004HydrogenPartition}\\
$\delta\_{HPE}$ & Partition coefficient for heat-producing elements & 0.001 & \citet{blundy2003PartitioningTrace}\\
\midrule
\multicolumn{4}{l}{\textit{Atmosphere \& escape}} \\[0.2em]
$A$ & Planetary albedo & 0.3&\\
$k_{0,\ce{H2}}$ & \ce{H2} absorption coefficient & \SI{2e-2}{\m\squared\per\kg}&after \citet{pierrehumbert2011HydrogenGreenhouse}\\
$k_{0,\ce{CO2}}$ & \ce{CO2} absorption coefficient & \SI{0.05}{\m\squared\per\kg} or \SI{0.001}{\m\squared\per\kg} & \citet{pujol2003AnalyticalInvestigation}\\
$k_{0,\ce{H2O}}$ & \ce{H2O} absorption coefficient & \SI{0.01}{\m\squared\per\kg}& \citet{abe1988EvolutionImpactGenerated}\\ 
$b_{\ce{CO2}}$ & Binary diffusion coefficient \ce{H2}-\ce{CO2} & \SI{3e21}{\per\m\per\s}& \citet{marrero1972GaseousDiffusion} \\
$b_{\ce{CO}}$ & Binary diffusion coefficient \ce{H2}-\ce{CO} & \SI{3e21}{\per\m\per\s}& \citet{marrero1972GaseousDiffusion}\\
$b_{\ce{H2O}}$ & Binary diffusion coefficient \ce{H2}-\ce{H2O} & \SI{4.3e21}{\per\m\per\s}& \citet{roberts1972MolecularDiffusion}\\
$\chi_0$ & Interpolation: \ce{H2} mixing ratio threshold & 0.15 \\
$w$ & Interpolation fall-off width & 0.01&\\
\midrule

\multicolumn{4}{l}{\textit{Carbon weathering}} \\[0.2em]
$X\_E$ & Present-day Earth mid-ocean ridge \ce{CO2} concentration in the melt & \SI{125}{ppm}& \citet{honing2019CarbonCycling}\\
$\xi\_E$ & Proportion of Earth's seafloor weathering to the total weathering rate & 0.15& \citet{foley2015RolePlate}\\
$f\_E$ & Present-day fraction of carbonates reaching Earth's mantle & 0.5& \citet{honing2019CarbonCycling}\\
$\phi_E$ & Fraction of stable carbonates during subduction & 0.8857& \citet{honing2019CarbonCycling}\\
$P_{\ce{CO2},\text{E}}$ & Present-day Earth atmospheric \ce{CO2} pressure & \SI{4e-4}{bar}& \citet{honing2019CarbonCycling}\\
$\alpha$ & Seafloor weathering scaling exponent & 0.23& \citet{foley2015RolePlate}\\
$A\_{decarb}$ & Decarbonation constant & \SI{3.125e-3}{\K\per\m}& \citet{foley2018CarbonCycling}\\
$B\_{decarb}$ & Decarbonation constant & \SI{835.5}{K}& \citet{foley2018CarbonCycling}\\
$E\_{bas}$ & Activation energy for basalt weathering & \SI{9.2e4}{\J\per\mol}& \citet{krissansen-totton2017ConstrainingClimate}\\
\end{tabularx}
\end{center}

\FloatBarrier
\newpage
\section{Additional figures}
\label{sec:additionl_figs}

\begin{figure}[ht!]
    \centering
    \includegraphics[width=0.7\linewidth]{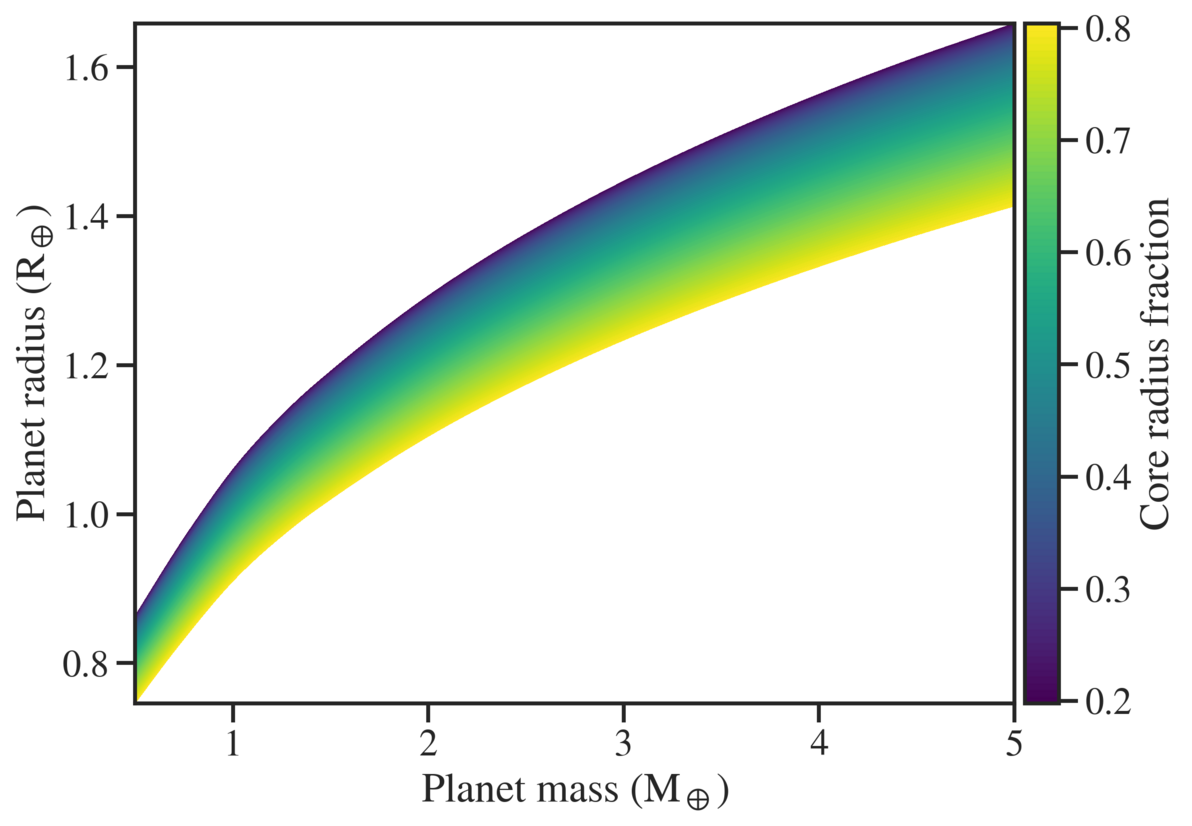}
    \caption{Mass-radius range of terrestrial planets considered in this study. The color map indicates the thickness of the iron core relative to the planet radius.}
    \label{fig:MRcrf}
\end{figure}

\begin{figure}[htb!]
    \centering
    \includegraphics[width=0.7\linewidth]{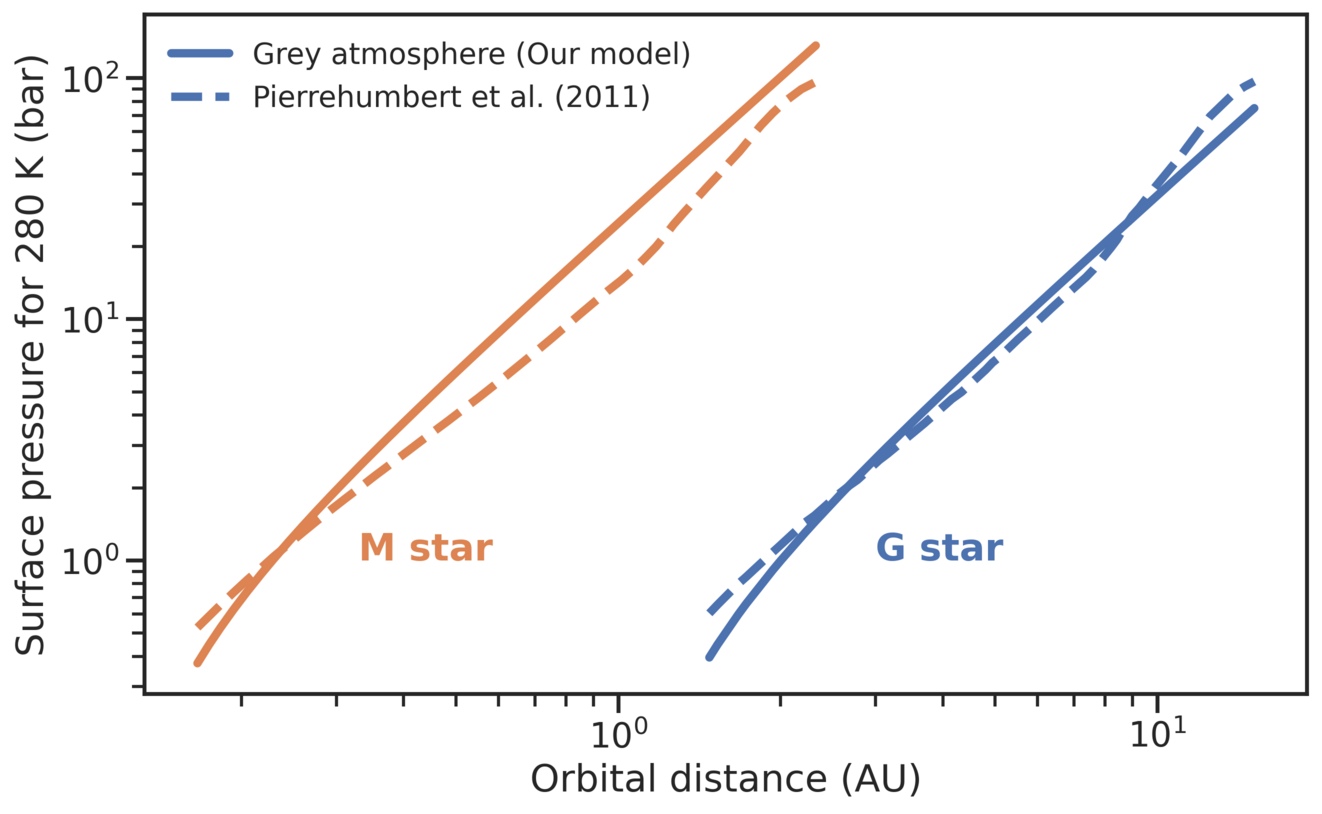}
    \caption{Amount of atmospheric \ce{H2} needed to keep the surface temperature at \SI{280}{K} as a function of orbital distance (after \citet{pierrehumbert2011HydrogenGreenhouse}). Dashed lines show the results from \citet{pierrehumbert2011HydrogenGreenhouse}, solid lines show the results of our atmosphere model, assuming a hydrogen absorption coefficient of $k_{0,\ce{H2}}$=\SI{2e-2}{\m\squared\per\kg}.}
    \label{fig:pierrehumbert}
\end{figure}

\begin{figure}[htb!]
    \centering
    \includegraphics[width=0.7\linewidth]{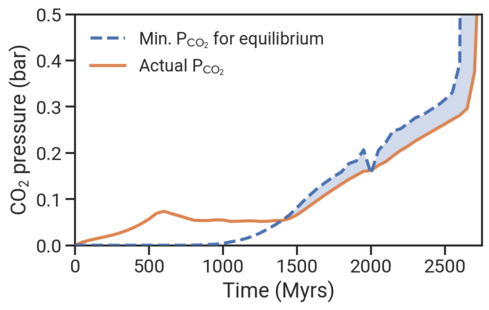}
    \caption{Example of limited weathering due to limited crustal growth. The blue dashed line shows the partial pressure of \ce{CO2} which would be necessary for weathering to be in equilibrium with the influx of \ce{CO2} from outgassing and decarbonation, based on the current crustal growth rate. The orange line shows the actual \ce{CO2} pressure in the atmosphere. The blue shaded region marks where the system is in disequilibrium, i.e. the influx of \ce{CO2} is larger than can be removed by weathering. This example correspond to the Earth-like model in Fig. \ref{fig:evolution}a-c.}
    \label{fig:supplylimit}
\end{figure}

\begin{figure}[htb!]
    \centering
    \includegraphics[width=0.7\linewidth]{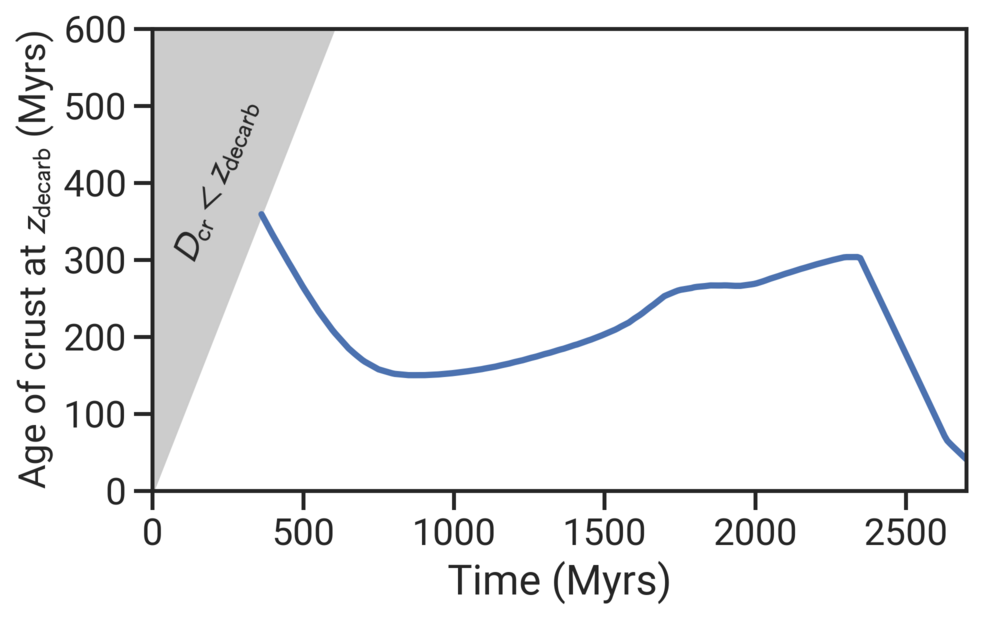}
    \caption{Example of the ``memory effect'' present in the stagnant-lid carbon cycle. The blue line shows the age of the crustal layers upon reaching the decarbonation depth $z\_{decarb}$, and therefore the time that has passed since the weathering rate was ``recorded'' into the crust. The shaded region marks where the crust is not yet old enough to reach the decarbonation depth. At \SI{2.3}{Gyrs}, the surface begins to heat up and thus younger and younger crust decarbonates as the decarbonation depth decreases. This example correspond to the Earth-like model shown in Fig. \ref{fig:evolution}a-c.}
    \label{fig:age_decarb}
\end{figure}

\begin{figure}[htb!]
    \centering
    \includegraphics[width=0.7\linewidth]{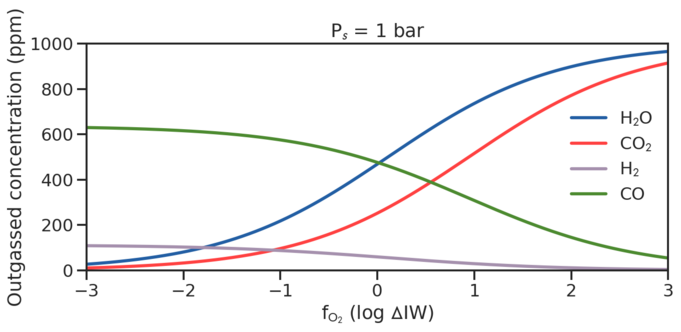}
    \caption{Example of the composition of outgassed species as a function of oxygen fugacity. Assumed here is a \SI{1}{bar} buffer atmosphere, and a volatile content in the melt of \SI{1000}{ppm} \ce{H2O} and \ce{CO2}, respectively, at a melt temperature of \SI{1500}{K}.}
    \label{fig:coh}
\end{figure}

\begin{figure}[htb!]
    \centering
    \includegraphics[width=\linewidth]{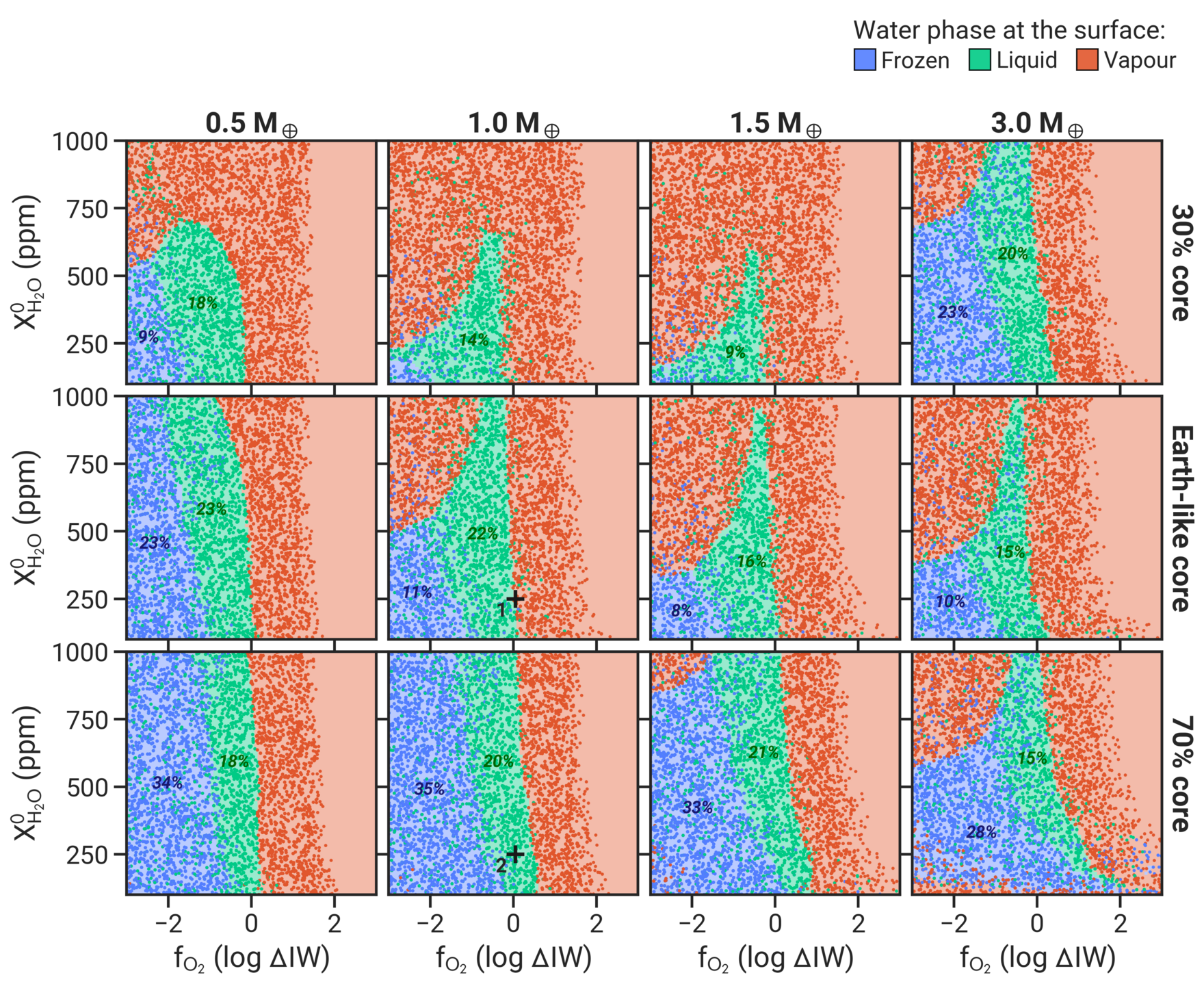}
    \caption{As Fig. \ref{fig:time_orbit}, but showing the prevailing surface conditions for water for different planet masses and core sizes.}
    \label{fig:mass_crf}
\end{figure}

\begin{figure}[htb!]
    \centering
    \includegraphics[width=0.9\linewidth]{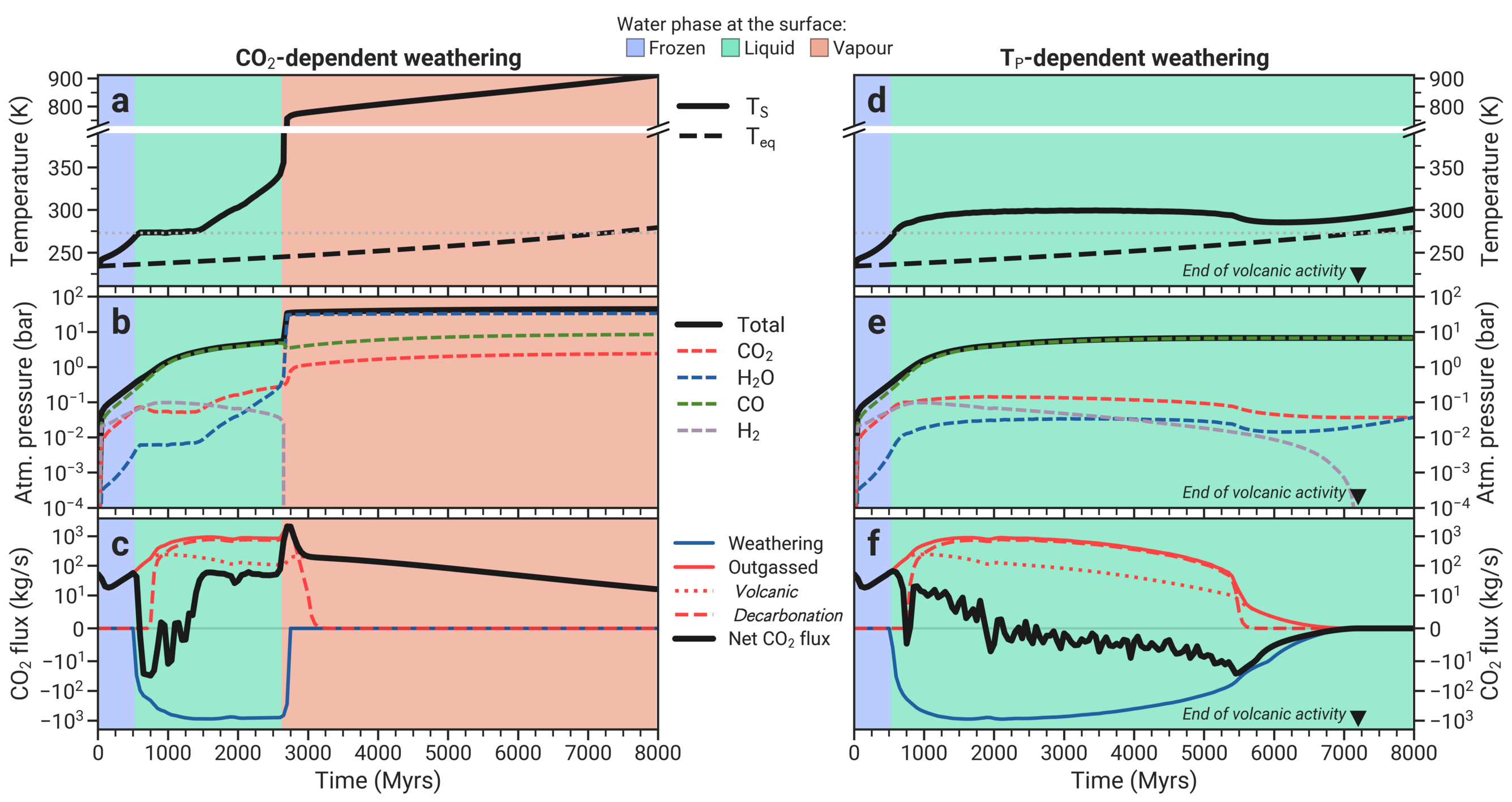}
    \caption{As in Fig.\ref{fig:evolution}, but showing a comparison between a \ce{CO2}-dependent (a, b, c) and a temperature-dependent weathering model (d, e, f). All other model parameters are the same as in Fig.\ref{fig:evolution}a-c.}
    \label{fig:evolution_weathering}
\end{figure}

\end{appendix}

\end{document}